\documentclass[journal,10pt]{IEEEtran}

\usepackage{ifpdf}
\usepackage{graphicx, color}
\usepackage{cite}
\usepackage{bm,bbm}
\usepackage{algorithm,algpseudocode}
\usepackage{booktabs}

\usepackage{url}

\usepackage{mathtools}

\usepackage{optidef}
\usepackage{multirow}

\ifCLASSOPTIONcaptionsoff
	\usepackage[nomarkers]{endfloat}
	\let\MYoriglatexcaption\caption
	\renewcommand{\caption}[2][\relax]{\MYoriglatexcaption[#2]{#2}}
\fi

\usepackage{amsmath,amssymb,amsfonts,mathtools,amsthm}

\DeclareMathOperator*{\argmax}{arg\,max}
\usepackage{comment}

\theoremstyle{definition}
\theoremstyle{definition}

\usepackage{subfigure}

\graphicspath{{./fig/}}

\begin{document}
\title{\fontsize{19}{25}\selectfont
  Millimeter Wave Communications on Overhead Messenger Wire:\\
  Deep Reinforcement Learning-Based Predictive Beam Tracking
}

\author{
  \normalsize Yusuke~Koda,~\IEEEmembership{Student~Member,~IEEE,}
  \normalsize Masao~Shinzaki,~\IEEEmembership{Student~Member,~IEEE,}
  \normalsize Koji~Yamamoto,~\IEEEmembership{Senior~Member,~IEEE,}
  \normalsize Takayuki~Nishio,~\IEEEmembership{Senior~Member,~IEEE,}
  \normalsize Masahiro~Morikura,~\IEEEmembership{Member,~IEEE,}
  \normalsize Yushi~Shirato, 
  \normalsize Daisei~Uchida, and
  \normalsize Naoki~Kita~\IEEEmembership{Member,~IEEE,}
  \thanks{
    Yusuke~Koda, Masao~Shinzaki, Koji~Yamamoto,  and Masahiro~Morikura are with the Graduate School of Informatics, Kyoto University, Yoshida-honmachi, Sakyo-ku, Kyoto 606-8501, Japan, e-mail: \{koda@imc.cce., shinzaki@imc.cce, kyamamot@, morikura@\}i.kyoto-u.ac.jp.

    Takayuki~Nishio is with School of Engineering, Tokyo Institute of Technology, Ookayama, Meguro-ku, Tokyo, 158-0084, Japan, e-mail: nishio@ict.e.titech.ac.jp.

    Yushi~Shirato, Daisei~Uchida, and Naoki~Kita are with the NTT Access Network Service Systems Laboratories, NTT Corporation, Yokosuka, 239-0847 Japan.
  }
}

\maketitle
\begin{abstract}
  This paper discusses the feasibility of beam tracking against dynamics in millimeter wave (mmWave) nodes placed on overhead messenger wires, including wind-forced perturbations and disturbances caused by impulsive forces to wires.
  Our main contribution is to answer whether or not historical positions and velocities of a mmWave node is useful to track directional beams given the complicated on-wire dynamics.
  To this end, we implement beam-tracking based on deep reinforcement learning (DRL) to learn the complicated relationships between the historical positions/velocities and appropriate beam steering angles.
  Our numerical evaluations yielded the following key insights:
  Against wind perturbations, an appropriate beam-tracking policy can be learned from the historical positions and velocities of a node.
  Meanwhile, against impulsive forces to the wire, the use of the position and velocity of the node is not necessarily sufficient owing to the rapid displacement of the node.
  To solve this, we propose to take advantage of the positional interaction on the wire by leveraging the positions/velocities of several points on the wire as state information in DRL.
  The results confirmed that this results in the avoidance of beam misalignment, which would not be possible by using only the position/velocity of the node.
\end{abstract}
\IEEEpeerreviewmaketitle



%
%

\section{Introduction}
\label{sec:introduction}
\IEEEPARstart{M}{illimeter-wave} (mmWave) communication technology presents great opportunities in a wide range of data-intensive applications.
Owing to the wider available bandwidth in the mmWave band, this greatly supports the 5G new radio (NR) access technology to compensate for the scarcity of spectrum in the microwave bands, thereby meeting the demand created by the increasing amount of mobile traffic\cite{niu2015survey, xiao2017millimeter}.
Apart from the 5G NR, the recent IEEE 802.11ay standard built upon the ratified IEEE 802.11ad standard proceeds beyond multi-gigabit-per-second connectivities to provide a data rate of 100\,Gbit/s\cite{chen2019millimeter, aldubaikhy2020mmwave,ghasempour2017ieee}.
This not only enables indoor high-speed Wi-Fi but also makes the concept of fiber-like connectivities such as outdoor mmWave wireless backhaul\cite{niu2015survey, rangan2014millimeter, ge20145g} or wireless-to-the-home systems\cite{aldubaikhy2020mmwave} a reality.

However, attempts to deploy mmWave nodes densely in outdoor environments could be thwarted by the limited available space for placing the nodes because of the line-of-sight (LOS) requirement.
In mmWave communications, a link budget is severely penalized by obstacles such as buildings and roadside trees\cite{rappaport2013millimeter, rappaport2015wideband}.
To prevent blockages by obstacles such as these and to reach the full potential of mmWave communications, it is important to position mmWave nodes such that obstacles do not obstruct the LOS between nodes that need to communicate.
Typically, owing to this LOS requirement, mmWave nodes would have to be placed at higher altitudes to prevent obstacles from obstructing LOS paths.
This limits the physical spaces for placing mmWave nodes to surfaces of buildings or telephone poles, and accordingly, the density of the mmWave nodes may depend on the density of suitably located buildings and telephone poles.

\begin{figure}[tb]
  \centering
  \includegraphics[width=0.9\columnwidth]{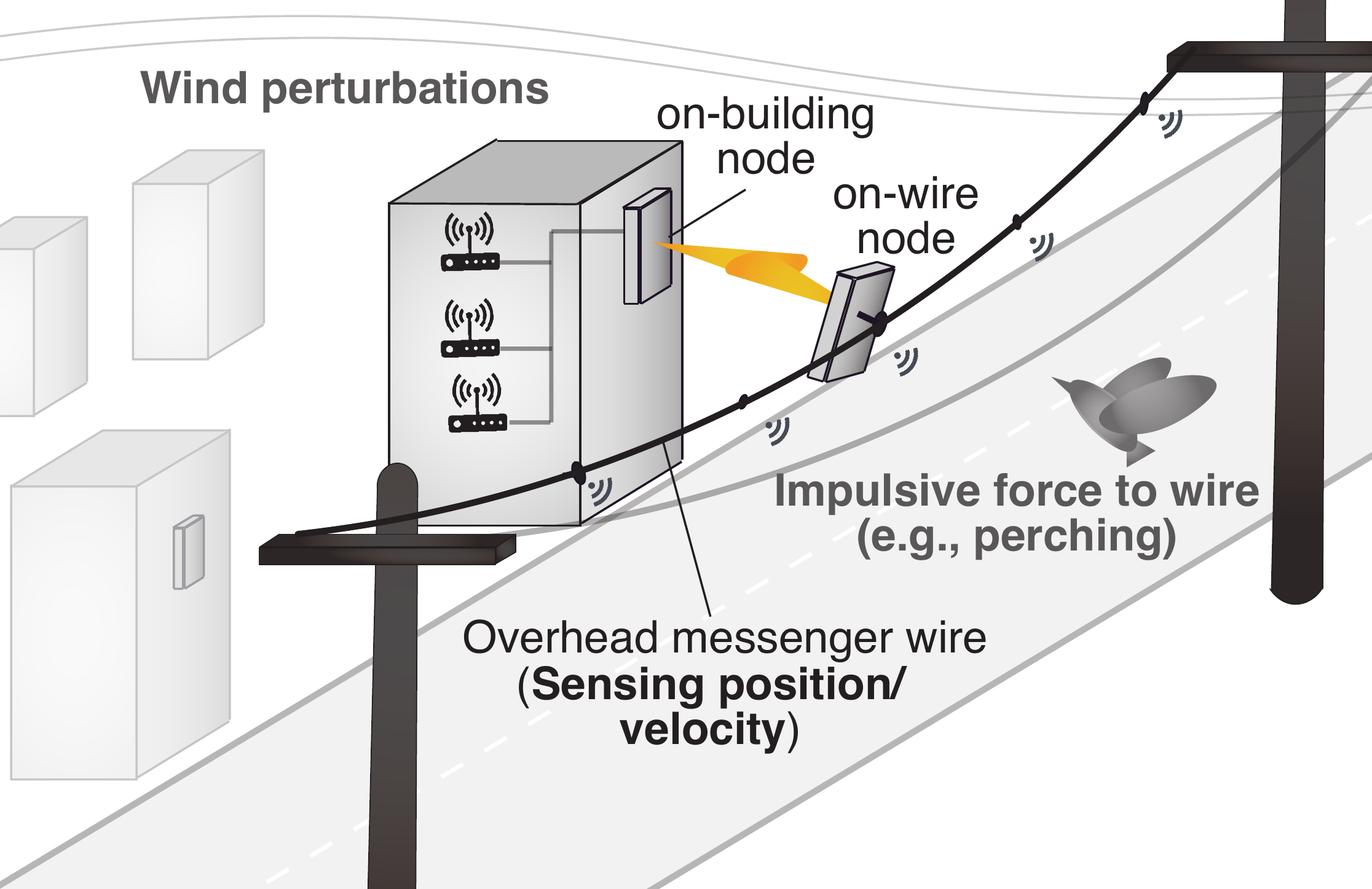}
  \caption{
    Scenario showing the on-wire deployment of mmWave nodes in a wireless-to-the-building use case.
    The beam should be tracked against various perturbations such as wind perturbations and other disturbances caused by impulsive forces to the wire.
    We investigate whether non-RF sensing information (e.g., position and velocity of several points on the wire) is useful or not to enhance the accuracy of the beam tracking against such perturbations specific to on-wire deployment.
  }
  \label{fig:system_model}
\end{figure}

To address this shortcoming, we explore the new possibility of installing mmWave nodes on overhead messenger wires as shown in Fig.~\ref{fig:system_model}.
The placement of mmWave nodes not only on buildings or on telephone poles but also on, for example, splice closures \textit{between telephone poles} offers flexibility with respect to the physical placement of nodes, thereby facilitating ensuring LOS connections.
In the course of this exploration, the key concern that remained is that on-wire deployment introduces dynamics in mmWave nodes owing to unprecedented root causes such as wind perturbations and other disturbances caused by impulsive forces to the messenger wire, which cause mmWave beam misalignment.
Beam misalignment is problematic because mmWave communications generally require directional antennas to compensate for the large path-loss in mmWave communications and to align the receiver and transmitter beams \cite{li2009spectrum}.
A naive approach to address beam misalignment is to let the transmitter and receiver scan their beam directions to determine the direction of maximum power periodically.
In a similar approach, in the IEEE 802.11ad standard, transmitter and receiver steering is decoupled using an omnidirectional antenna \cite{11ad, zhou2012efficient}.
Following the aforementioned on-wire dynamic conditions, search-based methods would require frequent beam-search to repair beam misalignment.
Consequently, the increases the overhead generated by the beam search, which deteriorates the spectral efficiency.

To combat the inefficiency resulting from the periodical beam-search, non-radio frequency (non-RF) information such as the positions and velocities of the mmWave nodes can be leveraged\cite{wang2018mmwave, wang2019mmwave, va2019online, arvinte2019beam, abdelreheem2016millimeter}.
However, acquisition of the positions and velocities is not necessarily synchronous to the beam-tracking operation, and, as a result, information about the current position of a mmWave node is not available.
In this case, we would have to predictively determine an appropriate beam steering angle while referring to the past positions of mmWave nodes.
In the case of the on-wire deployment of a mmWave node, our key question is:
\textbf{In view of the complicated perturbations in the on-wire deployment, would past information about the position and velocity of a mmWave node be useful to feasibly track directional beams?}

This work attempts to answer this question by proposing a feasible system for tracking directional beams based on historical values of the position and velocity under an on-wire mmWave node deployment.
Specifically, for the examination of the feasibility, we leverage deep reinforcement learning (DRL) to determine beam steering angles based on past position/velocity information of a mmWave node. 
The reason for the choice of DRL is discussed in Sections~\ref{subsec:contribution} and \ref{subsec:problem_formulation} in this paper. 
It should be noted that many studies have applied DRL to mmWave communication\cite{khan2019reinforcement, feng2019dealing, mismar2019deep, wang2020precodernet}, among which one of the studies \cite{mismar2019deep} already proposed DRL-based mmWave beam tracking, yet considered a simpler scenario of on-ground mobile environments.
However, this emergence of DRL is precisely why it is important to proceed beyond simple on-ground mobile environments and to explore the feasibility of targeting more complicated dynamic environments such as those in which mmWave nodes are placed on messenger wires.

Importantly, another benefit of examining the aforementioned feasibility is that it is expected to shed light on mmWave node deployments in other unstable surroundings.
These surroundings are exemplified by lampposts and traffic lights.
These locations are subject to perturbations similar to those of messenger wires such as swaying due to wind and rapid displacements due to forces applied to the poles caused by airborne objects or ground vibrations from road traffics.
The deployment of mmWave nodes in such places greatly supports not only the wireless-to-the-building use cases in Fig.~\ref{fig:system_model}, but also, though not limited to, upcoming mmWave vehicle-to-everything (V2X)\cite{choi2016millimeter, sakaguchi2020towards}.
This is because such placements allow vehicles to communicate with mmWave nodes regarded as roadside units in the proximity of them, enabling to exchange large volume sensory data crucial to support autonomous driving (e.g., camera images, light detection and ranging data, or other radar data)\cite{sakaguchi2020towards}.
Hence, via examining the beam tracking in an on-wire mmWave node placement, this work attempts to open up the opportunities to deploy mmWave nodes in various unstable placements towards diverse mmWave use cases.

\subsection{Related Work}
MmWave communications necessitate the use of directional antennas to compensate for the large path loss in the mmWave spectrum.
Accordingly, considering the dynamics of mmWave nodes, beam tracking is required to maintain the received power; consequently, the design of efficient beam tracking methods has attracted considerable research interest.
This section provides an overview of previous work and discusses the differences among the proposed solutions.

\vspace{.5em}\noindent \textbf{Search-based Beam Tracking.}\quad
To perform beam tracking in mobile environments, a typical approach entails periodically searching for an appropriate beam pair between transmitters and receivers.
The most naive approach is to exhaustively scan all possible beam pairs during the beam search.
In the IEEE 802.11ad standard, transmitter and receiver steering is decoupled using an omnidirectional antenna\cite{11ad}, which is more efficient than exhaustive beam search.
More efficient methods capable of further reducing the search spaces are exemplified by subspace sampling in hierarchial codebooks\cite{hur2013millimeter, li2012efficient, tsang2011coding}, multi-stage search \cite{wang2009beam}, agile-link leveraging multi-armed beams\cite{hassanieh2018fast}, parallel-adaptive beam training\cite{de2017millimeter}, the use of location information\cite{abdelreheem2016millimeter}, and hybrid beam forming with compressing sensing\cite{alkhateeb2014channel}.
In general, this type of approach requires channel observation of the scanned beam pairs, which generates overhead proportionate to the number of beam pairs in terms of spectrum utilization.
Unlike these solutions, our approach does not require beam scanning by learning appropriate beam angles from past experience and applying the beam-tracking policy afterward.


\vspace{.5em}\noindent \textbf{Learning-based Beam Tracking.}\quad
To avoid overhead in terms of spectrum utilization, several studies proposed learning-based beam tracking.
Beam scanning procedures can be eliminated by learning an appropriate beam-tracking policy from past experiences and applying the learned policy afterward.
In this approach, the design of input information for learned machine learning (ML) models to be associated with beam steering angles is divided into two categories: RF information or non-RF information.
\begin{enumerate}
  \item \textbf{RF Information.}\quad
        Many studies devoted to developing ML models to determine the appropriate beam steering angles leveraged RF information such as the channel state information or signal-to-noise ratio as the input\cite{alkhateeb2018deep, zhou2018deep, araujo2019beam, chang2019learning, jeong2020online}.
        In \cite{alkhateeb2018deep}, beam tracking based on supervised learning was proposed to learn achievable data rates corresponding to possible beam pairs while accepting input information of pilot signals in coordinated BSs.
        Once the relationship between the data rates of possible beam pairs and pilot signals is learned, appropriate beam pairs can be determined without scanning possible beams.
        In \cite{zhou2018deep}, multiple transmitter/receiver pairs were considered, wherein a beam-tracking policy that avoids interference was learned.
        An RL-based approach was also proposed\cite{araujo2019beam, jeong2020online}, wherein a channel observation and current beam pair are regarded as a state, and the beam steering angle is regarded as an action.
        In \cite{jeong2020online}, the beam steering angle and beam width were jointly optimized via DRL.
        However, this approach still requires channel observations to obtain RF information to determine the steering angle, which may incur overhead particularly in a mobile environment in which the beam frequently needs to be fixed.
        Unlike these studies, we leverage non-RF information as input to our ML models.
  \item \vspace{.3em}\textbf{Non-RF Information.}\quad
        Leveraging non-RF information as the input enables the elimination of channel observation to determine the beam steering angles.
        Supervised learning-based beam tracking that learns the relationship between user locations and appropriate beam steering angles was proposed \cite{arvinte2019beam}.
        In another study, DRL was proposed by regarding the positions of mobile users as a state, which is the input of the ML model \cite{mismar2019deep}.
        The positions of cars were utilized as state information in RL-based beam tracking \cite{klautau20185g}.
        In addition to the tracked vehicles, the positions of surrounding vehicles were also utilized to determine an appropriate path \cite{wang2018mmwave, wang2019mmwave, va2019online}.
        In recent work \cite{alrabeiah2020millimeter}, visual information was leveraged to monitor vehicles.
        However, this assumes simpler two-dimensional on-ground dynamics of tracked mobile devices and the availability of the current positions of the devices.
        Contrary to these studies, we investigate the feasibility of learning under more complicated three-dimensional on-wire dynamics in a mmWave node while relaxing the assumption that synchronous non-RF information is acquired.
\end{enumerate}

To summarize, the main difference between this work and reports in the literature is \textit{1)} the consideration of dynamics specific to on-wire deployments and \textit{2)} using only historical information about the position and velocity of a mmWave node.
In on-wire deployment, the dynamics of a mmWave node essentially differ from those of on-ground mobile environments in the sense that the dynamics include oscillations due to the tensile force of wires and rapid position validations as a result of impulsive forces to wires.
Hence, the literature does not provide straightforward answers as to whether an appropriate beam-tracking policy can be feasibly learned based on past information about the position and velocity.

\subsection{Contributions and Organization of This Paper}
\label{subsec:contribution}
The salient contributions of this paper are summarized as follows:
\begin{itemize}
  \item 
   We demonstrate the feasibility of tracking directional beams of a mmWave node placed on an overhead messenger wire and that is subject to complicated on-wire dynamics, which is an unexplored scenario in the wireless communications research area. 
  Specifically, motivated by the aforementioned question, we examine whether DRL could be leveraged to predictively determine beam steering angles based on past information about the position and velocity of the mmWave node. 
  The choice of DRL is because of its powerful capability to learn complicated relationships between state information (i.e., historical information about the position and velocity of the mmWave node in our case) and the appropriate actions (i.e., beam steering angles) by using neural networks (NNs).
  Although DRL has previously been used for beam tracking \cite{mismar2019deep}, to the best of our knowledge, the feasibility of tracking beams under on-wire dynamics based on past information about the position and velocity has not yet been investigated.
  \item 
        Examining the feasibility of the beam tracking in view of impulsive forces to a messenger wire, we find it is useful to take advantage of the positional interaction among several points on the wire, which are specific characteristics of on-wire deployments.
        More specifically, as state information in DRL, we leverage the information pertaining to the position and velocity of several points on the wire and demonstrate that it is possible to avoid beam misalignment that could not be avoided based only on the position/velocity of the mmWave node.
        This is because the variation in the position of points at which the impulsive force applied propagates to the mmWave node via the tensile force.
        Hence, the positions/velocities of several points on the wire not only provide information about the position of the mmWave node, they also indicate the way in which the position of the mmWave node could be expected to vary in the future, leading to more accurate beam tracking.
\end{itemize}
It should be noted that the main objective of this work is to answer the aforementioned question, which is again stated here: Is historical information about the position and velocity of mmWave nodes useful to feasibly track directional beams under perturbations specific to on-wire deployments?
To focus on this, we assume that past information about the position and velocity of a mmWave node is available beforehand, and we consider an in-depth discussion of the acquisition of such information to be beyond the scope of this paper.
This gives us the aforementioned insight, which is sufficiently helpful to explore the possibility of on-wire mmWave node placements.
In addition, compared with the preliminary version of this work presented at IEEE VTC-fall 2020\cite{shinzaki2020deep}, this work addresses the aforementioned question more comprehensively by investigating various characteristics of messenger wires.
Moreover, this work considers not only wind perturbations, but also impulsive forces to wires, whereas the preliminary version in \cite{shinzaki2020deep} considered only wind perturbations.
Accordingly, this work examines the usage of multiple position/velocity sensors to combat against the disturbances caused by impulsive forces, and this insight was not provided in the previous work.

The remainder of this paper is organized as follows.
In Section \ref{sec:system_model_problem_formulation}, we present the system model and problem formulation.
In Section \ref{sec:DeepRL}, we provide the DRL-based beam-tracking framework to solve the formulated problem.
In Section \ref{sec:simulation}, we describe the numerical evaluation of the DRL-based beam-tracking framework by reproducing the on-wire dynamics subject to the aforementioned perturbations.
Finally, we provide our concluding remarks in Section \ref{sec:conclusion}.

%
%
\section{System Model and Problem Formulation}
\label{sec:system_model_problem_formulation}

\subsection{System Model}
\label{subsec:system_model}

\noindent\textbf{On-wire mmWave Node Deployment.}\quad
We consider an on-wire mmWave node deployment wherein a single mmWave node serves another node mounted on a vertical building surface as shown in Fig.~\ref{fig:system_model}.
The former mmWave node is attached to the messenger wire, with its placement being motivated by the necessity to ensure a LOS connection.
The on-wire mmWave node relays data from the node attached to the building, forwarding indoor data traffic into core networks via the messenger wire, and vice-versa.
The position of the on-wire node is perturbed by wind and impulsive forces to the messenger wire due to airborne objects exemplified by birds and unmanned aerial vehicles.
Against these perturbations, the mmWave directional beam is periodically steered by leveraging non-RF information exchanged in the messenger wire.

To investigate the way in which non-RF information aids beam tracking, we consider that, as an example, the positions and velocities at certain points on the messenger wire are available with multiple sensors.
These sensors obtain the positions and velocities in an absolute coordinates in an error-free manner with constant intervals.
This assumption is motivated by our focus on the perturbations specific to the on-wire, and hence, we eliminate other root causes of beam-misalignments such as sensory errors.
Meanwhile, motivated by the question in Section~\ref{sec:introduction}, the acquisition of the sensor information does not need to be synchronous to the beam-tracking operations because of, for example, inherent sensor delay or the delay for transmitting the sensory information to beam-tracking agents.
To ensure the generality, we do not specify the root cause of the time-difference between the sensory acquisition and the beam-tracking operations by just denoting the difference as $T$, which is termed look-back time and is detailed in Section~\ref{subsec:problem_formulation}.


\vspace{.5em}\noindent \textbf{Dynamics of Overhead Messenger Wire.}\quad
Points on messenger wire interact with adjacent ones via tensile force.
To imitate such positional interaction, we consider the messenger wire as a chain of points wherein adjacent points are interconnected with a spring\cite{maor1975discrete}.
Specifically, the messenger wire is considered as $N$ material points (see. Fig.~\ref{fig:dynamics_wire} in the next section) with the mass of $m/N$, where the adjacent points are physically connected by springs with spring constant $k_0$.
Note that $m$ denotes the mass of the entire messenger wire.
The considered $N$ points encompass two endpoints of the messenger wire, and we term the endpoints $\mathrm{P}1$ and $\mathrm{P}N$.
The other points are termed $\mathrm{P2}, \dots, \mathrm{P}N - 1$ in the order of proximity from $\mathrm{P}1$.

To obtain the position of $\mathrm{P}1, \dots, \mathrm{P}N$ affected by the aforementioned interaction with proximity points, wind perturbations, and impulsive forces, we consider the following dynamics model.
For $i = 1, 2, \dots, N$, let $\bm{a}_i(t), \bm v_{i}(t), \bm x_i(t)\in\mathbb{R}^3$ denote the acceleration, velocity, and position of $\mathrm{P}i$ in a three-dimensional Euclidean space at time instant $t$, respectively.
Without loss of generality, we assume that the origin $o$ is located at the mid-point of the messenger wire under the stable status without wind perturbations and impulsive forces.
The components of the acceleration $\bm{a}_i(t)$ are denoted by $a_{i, \mathrm{X}}(t), a_{i, \mathrm{Y}}(t), a_{i, \mathrm{Z}}(t)\in\mathbb{R}$, where the superscript $\mathrm{X}$ and $\mathrm{Y}$ denote the horizontal component, and $\mathrm{Z}$ denotes the vertical component.
This notation is also applied to $\bm v_{i}(t), \bm x_i(t)$.
The endpoints $\mathrm{P}1$ and $\mathrm{P}N$ are fixed, and hence $\bm a_1(t) = \bm a_N(t) = \bm v_1(t) = \bm v_N(t) = \bm 0$.
For $i=2,3,\dots,N-1$, $\bm a_i(t)$ modeled as:
\begin{multline}
  \label{eq:dynamics_acceleration}
  \bm a_i(t) = \bm g + \underbrace{\frac{k_0N}{m}\bigl[\bm x_{i+1}(t) + \bm x_{i-1}(t) - 2\bm x_i(t)\bigr]}_{\text{From tensile force yielding positional interaction}}
  \\ + \underbrace{\frac{N}{m} \bm{F}_i(t)}_{\substack{\text{From impulsive force}}},
\end{multline}
where $\bm g\in\mathbb R^3$ denotes the gravitational acceleration.
As an example of wind dynamics, we consider an Ornstein-Uhlenbeck process with an average wind velocity $\bm v_0(t)\in\mathbb{R}^3$\cite{shiri2019massive}.
Given the above acceleration, the derivatives of velocity and position $\mathrm d\bm v_i(t)$ and $\mathrm d\bm r_i(t)$ for $i=2,3,\dots,N-1$ are given by:
\begin{align}
  \mathrm d\bm v_i(t) & = \bm a_i(t)\,\mathrm dt \underbrace{-c_0\bigl[\bm v_i(t)-\bm v_\mathrm o(t)\bigr]\,\mathrm dt + \bm V_\mathrm o \mathrm d\bm W_i(t)}_{\text{From wind perturbation}}, \label{eq:dynamics_velocity} \\
  \mathrm d\bm x_i(t) & = \bm v_i(t)\,\mathrm dt, \label{eq:dynamics_position}
\end{align}
where $c_0$, $\bm V_\mathrm o\in\mathbb R^{3\times 3}$, $\bm W_i(t)\in\mathbb R^3$ denote the drag constant, the covariance matrix of the wind speed, and the standard Wiener process, respectively.

\vspace{.5em}\noindent\textbf{Radio Wave Propagation and Directional Antenna Model.}\quad
Owing to the placement to ensure LOS connectivity between on-wire and on-building nodes, we assume that the the direct waves are dominant.
Accordingly, we only consider the path loss in the channel model.
Given this, the received power denoted as $P_{\mathrm{RX}}$ is given by:
\begin{align}
  P_{\mathrm{RX}} =  P_\mathrm{TX} G_\mathrm{TX}G_\mathrm{RX}\beta r^{-\alpha},
  \label{eq:channel}
\end{align}
where $P_\mathrm{TX}$, $\beta$, $\alpha$ $r$ denote the transmit power, path loss at the unit distance, path loss exponent, and distance between the on-wire and on-building nodes, respectively.
For subscript $j\in\{\mathrm{TX}, \mathrm{RX}\}$, $G_j$ denotes the transmitter/receiver antenna gain, which is discussed next in detail.

Because we consider the three-dimensional dynamics in the on-wire mmWave node, we model the antenna radiation pattern following the 3GPP antenna radiation model with three-dimensional uniform planar antennas\cite{3gpp}.
Following the model, the antenna radiation pattern $A_j(\theta,\phi, \theta_{\mathrm{s}, j}, \phi_{\mathrm{s}, j})$ is given by
\begin{align}
  \label{eq:antenna_gain}
  A_j(\theta, \phi, \theta_{\mathrm{s}, j}, \phi_{\mathrm{s}, j}) = A_\mathrm{E}(\theta, \phi) + \mathit{AF}(\theta, \phi, \theta_{\mathrm{s}, j}, \phi_{\mathrm{s}, j}),
\end{align}
where $(\theta, \phi)$ are the zenith and azimuth angles relative to  reference angles $(\theta_{\mathrm{s}, j}, \phi_{\mathrm{s}, j})$ exhibiting the maximum directivity gain, which is termed main lobe steering direction.
In \eqref{eq:antenna_gain}, $A_\mathrm{E}(\theta, \phi)$ and $\mathit{AF}(\theta, \phi, \theta_{\mathrm{s}, j}, \phi_{\mathrm{s}, j})$ denote the element radiation pattern and the array factor, respectively, which are precisely given in \cite{3gpp}.
It should be noted that $G_\mathrm{TX}$ and $G_{\mathrm{RX}}$ in \eqref{eq:channel} are given by substituting the angles of arrival and angles of departure into $A_{\mathrm{TX}}(\theta, \phi, \theta_{\mathrm{s}, \mathrm{TX}}, \phi_{\mathrm{s}, \mathrm{TX}})$ and $A_{\mathrm{RX}}(\theta, \phi, \theta_{\mathrm{s}, \mathrm{RX}}, \phi_{\mathrm{s}, \mathrm{RX}})$, respectively.
More specifically, $G_\mathrm{TX}$ and $G_{\mathrm{RX}}$ are given by:
\begin{align}
   & G_{\mathrm{TX}} = A_\mathrm{TX}(\theta_{\mathrm{AoD}}, \phi_{\mathrm{AoD}}, \theta_{\mathrm{s}, \mathrm{TX}}, \phi_{\mathrm{s}, \mathrm{TX}}),  \\
   & G_{\mathrm{RX}} = A_\mathrm{RX}(\theta_{\mathrm{AoA}}, \phi_{\mathrm{AoA}}, \theta_{\mathrm{s}, \mathrm{RX}}, \phi_{\mathrm{s} , \mathrm{RX}}),
\end{align}
where $(\theta_{\mathrm{AoD}}, \phi_{\mathrm{AoD}})$ and $(\theta_{\mathrm{AoA}}, \phi_{\mathrm{AoA}})$ denote the angle of departure and angle of arrival relative to the aforementioned reference angles, respectively.

Beam tracking is generally performed by configuring the main lobe steering direction $(\theta_{\mathrm{s}, j}, \phi_{\mathrm{s}, j})$ in the array factor, and hence, we note the mathematical expression of the array factor in detail.
For the sake of simplicity, we omit the subscript $j$ indicating transmitter or receiver.
The array factor for an array of $n=n_{\mathrm{v}}\times n_{\mathrm{h}}$ elements is expressed as
\begin{align}
  \mathit{AF}(\theta, \phi, \theta_{\mathrm{s}}, \phi_{\mathrm{s}}) = 10\log_{10}\Bigl[1+\rho\Bigl(\left|\bm a\cdot\bm w^{\mathrm{T}}\right|^{2}-1\Bigr)\Bigr],
  \label{eq:array_factor}
\end{align}
where $n_{\mathrm{v}}$, $n_{\mathrm{h}}$, $\rho$ denote the number of vertical elements, the number of horizontal elements, and correlation coefficient, respectively.
In \eqref{eq:array_factor}, $\bm a\in\mathbb C^{n_{\mathrm{v}}n_{\mathrm{h}}}$ and $\bm w\in\mathbb C^{n_{\mathrm{v}}n_{\mathrm{h}}}$ denote the amplitude vector and beamforming vector, respectively, and all elements of $\bm a$ are generally set as $1/n_{\mathrm{v}}n_{\mathrm{h}}$.
Hence, the problem of tracking beams falls into setting the beamforming vector $\bm w$ such that an appropriate main lobe steering direction $(\theta_{\mathrm{s}}, \phi_{\mathrm{s}})$ is configured.
The relationship between $\bm w$ and $(\theta_{\mathrm{s}}, \phi_{\mathrm{s}})$ is given by
\begin{align*}
  \bm w   & = \left[w_{1,1}, w_{1,2}, \ldots, w_{n_{\mathrm{v}}, n_{\mathrm{h}}}\right]^{\mathrm{T}}, \notag                       \\
  w_{p,r} & = \mathrm e^{\mathrm j2\pi[(p-1)\Delta_{\mathrm{V}}\Psi_p/\lambda + (r-1)\Delta_{\mathrm{H}}\Psi_r]/\lambda}, \notag                         \\
  \Psi_p  & = \cos(\theta + \theta_{\mathrm{s}}) - \cos \theta_{\mathrm{s}}, \notag                                                \\
  \Psi_r  & = \sin(\theta + \theta_{\mathrm{s}})\sin(\phi + \phi_{\mathrm{s}}) - \sin\theta_{\mathrm{s}}\sin\phi_{\mathrm{s}},
\end{align*}
where $\Delta_{\mathrm{V}}$ and $\Delta_{\mathrm{H}}$ denote the spacing distances between the vertical and horizontal elements of the uniform planar array, respectively.

\subsection{Problem Formulation}
\label{subsec:problem_formulation}
Our objective is to obtain a strategy for refining the beam direction against the aforementioned perturbations.
Without loss of generality, we only consider the beam tracking of the on-wire node by employing isotropic antennas in the other mmWave node, i.e., $G_{\mathrm{RX}} = 0$\,dBi over every angle.
We consider the problem of sequentially determining whether the main lobe steering direction is varied with a predefined angle difference $A$.
Considering a finite time duration $t\in [0, T_{\mathrm{d}}]$, where $T_{\mathrm{d}}\in\mathbb{R}$, we let superscript $(k)$ denote the time step, where $k \coloneqq \lfloor t/\tau \rfloor$, and $\tau$ is the period for which beam refinement is performed.
Given this, for a sufficiently large integer $K_{\mathrm{d}} \coloneqq \lfloor T_{\mathrm{d}}/\tau \rfloor$, the objective problem is formulated as follows:
\begin{maxi!}[1]
{\substack{\bigl(a_{\theta}^{(k)}\!, \ a_{\phi}^{(k)}\bigr)}}
{\sum_{k = 0}^{K_{\mathrm{d}}} P^{(k)}_{\mathrm{RX}}}{}{}
\addConstraint{\eqref{eq:dynamics_acceleration},}{\eqref{eq:dynamics_velocity},\eqref{eq:dynamics_position}}{}\label{eq_const:dynamics}
\addConstraint{r^{(k)}}{=\left\Vert\bm x_{i_{\mathrm{TX}}}(k\tau) - \bm x_{\mathrm{RX}}\right\Vert}{}
\addConstraint{\theta^{(k)}_{\mathrm{AoD}}}{=\arccos \frac{ x_{\mathrm{RX}, \mathrm{Z}} -  x_{i_{\mathrm{TX}},\mathrm{Z}}(k\tau)}{r^{(k)}}} - \theta^{(k)}_{\mathrm{s}, \mathrm{TX}}{} \label{eq_const:AOItheta}
\addConstraint{\phi^{(k)}_{\mathrm{AoD}}}{=\arctan \frac{x_{\mathrm{RX}, \mathrm{Y}} - x_{i_{\mathrm{TX}}, \mathrm{Y}}(k\tau)}{x_{\mathrm{RX}, \mathrm{X}} - x_{i_{\mathrm{TX}}, \mathrm{X}}(k\tau)}} - \phi^{(k)}_{\mathrm{s}, \mathrm{TX}}{}\label{eq_const:AOIphi}
\addConstraint{\theta^{(k)}_{\mathrm{s}, \mathrm{TX}}}{=\,\theta^{(k - 1)}_{\mathrm{s}, \mathrm{TX}}+\Delta_{\theta}^{(k)}}{}
\addConstraint{\phi^{(k)}_{\mathrm{s}, \mathrm{TX}}}{=\,\phi^{(k - 1)}_{\mathrm{s}, \mathrm{TX}}+\Delta_{\phi}^{(k)}}{}
\addConstraint{\Delta_{\theta}^{(k)}, \Delta_{\phi}^{(k)}}{\in\,\{-A, A, 0\}}{}.\label{eq_const:action}
\end{maxi!}
In the above problem, $i_{\mathrm{TX}}\in\{1, 2, \dots, N\}$ denotes the point at which the on-wire node is attached, and $\bm x_{\mathrm{RX}} = [x_{\mathrm{RX}, \mathrm{X}}, x_{\mathrm{RX}, \mathrm{Y}}, x_{\mathrm{RX}, \mathrm{Z}}]^\mathrm{T}$ is the position at which the fixed node is mounted.
The constraint \eqref{eq_const:dynamics} determines the dynamics of the messenger wire and the attached on-wire node affected by wind perturbations and impulsive forces, which is an essential difference from the problem formulations in the literature regarding beam-tracking methods.
The constraints \eqref{eq_const:AOItheta} and \eqref{eq_const:AOIphi} determine the effect of the dynamics of the messenger wire and main lobe steering angles on the dynamics of the relative angles of departure, which can be obtained by a simple geometric calculation.

The above problem can be easily solved by employing delay-free position information, that is, when determining the main lobe steering direction at each time instant $k\tau$ for $k = 1, 2, \dots$, the positions of the on-wire node at the same time instant, i.e., $\bm x_{i_{\mathrm{TX}}}(k\tau)$, are available.
Indeed, given this, the optimal strategy is to configure the main lobe steering direction such that the main lobe points at the fixed node.
However, as discussed above, we concern that it may not be easy to obtain such perfect position information in terms of time because of the processing delay or the transmission delay of the information or the difference between the interval of beam tracking and that of the position acquisition.
In this case, it is necessary to \textit{look back} at past positions of the on-wire node and determine an appropriate main lobe steering direction while predicting the variation in the on-wire node position within the look-back time.
Let the look-back time be denoted by $T$ hereinafter.

This motivated us to implement deep learning-based beam tracking.
Deep learning successfully builds prediction models in a data-driven manner by leveraging NNs, even when the relationship between inputs and outputs cannot be given in an explicit form owing to its complexity.
Hence, applying deep learning to our problem, we learn an appropriate relationship between past look-back positions of the on-wire node and beam direction even when, within the look-back time $T$, the position of the node is subject to complicated on-wire dynamics.
The approach we employ for deep learning-based beam tracking is to use deep RL to find the optimal beam-tracking policy based on the look-back position information on the on-wire node.
In the next section, we detail the DRL-based beam tracking that is implemented in this work.


%
%
\section{Implemented DRL-Based Beam Tracking}
\label{sec:DeepRL}
DRL learns an NN model that relates observable information in a considered environment termed the \textit{state} to an appropriate \textit{action} that maximizes a long-term \textit{reward}.
In this section, we detail the state, action, and reward we used to solve the aforementioned beam-tracking problem.
Subsequently, we discuss the importance of leveraging not only the information about the on-wire node but also information about other points on the messenger wire, focusing on the positional interaction specific to the on-wire deployment.

\subsection{State, Action, and Reward}
\vspace{.3em}\noindent \textbf{State.}\quad
The state is defined as the look-back positions and velocities of the on-wire node and main lobe steering direction.
More specifically, the state at each time instant $s_k$ for $k\in\mathbb{N}$ includes:
\begin{itemize}
  \item look-back positions of the on-wire node, i.e., $\bm{x}_{i_{\mathrm{TX}}}(k\tau - T)$.
  \item look-back velocities of the on-wire node, i.e., $\bm{v}_{i_{\mathrm{TX}}}(k\tau - T)$.
  \item The current main lobe steering direction $\bm b^{(k)}$, which is given by
        \begin{multline}
          \label{eq:beam_direction}
          \bm b^{(k)} \coloneqq
          \\
          \left[\sin\theta_{\mathrm{s}, \mathrm{TX}}^{(k)}\cos\phi_{\mathrm{s}, \mathrm{TX}}^{(k)}, \sin\theta_{\mathrm{s}, \mathrm{TX}}^{(k)}\sin\phi_{\mathrm{s}, \mathrm{TX}}^{(k)}, \cos\theta_{\mathrm{s}, \mathrm{TX}}^{(k)}\right]^{\mathrm{T}}.
        \end{multline}
\end{itemize}

\vspace{.3em}\noindent \textbf{Action.}\quad
The action determines whether the main lobe is steered at a pre-defined angle $A$ or not.
Specifically, each action at time instant $k\tau$ for $k\in\mathbb{N}$ is defined as:
\begin{align}
  \mathit{action}_k\coloneqq \Bigl(\Delta_{\theta}^{(k)}, \Delta_{\phi}^{(k)}\Bigr),
\end{align}
where the $\Delta_{\theta}^{(k)}$ and $\Delta_{\phi}^{(k)}$ correspond to the optimized direction of the main lobe in the aforementioned problem and are subject to the constraint \eqref{eq_const:action}.

\vspace{.3em}\noindent\textbf{Reward.}\quad
The immediate reward is defined as the instantaneous received signal power.
The reward at each time instant $r_k$ is defined as
\begin{align}
  r_k \coloneqq P_{\mathrm{RX}}^{(k)}.
\end{align}
However, applying the above reward definition may result in an unstable learning performance owing to the large variation of the reward values.
Hence, following a previous study \cite{mnih2015human}, we use a proxy reward that is scaled via clipping.
Specifically, the proxy reward $\hat{r}_k$ is given by
\begin{align}
  \label{eq:clipping}
  \hat{r}_k =
  \begin{cases}
    1,                             & (r_k-b_\mathrm c)/d_\mathrm c >1;          \\
    (r_k-b_\mathrm c)/d_\mathrm c, & -1\leq(r_k-b_\mathrm c)/d_\mathrm c\leq 1; \\
    -1,                            & (r_k-b_\mathrm c)/d_\mathrm c < -1,
  \end{cases}
\end{align}
where $b, d_{\mathrm{c}}\in\mathbb{R}$ denote the offset and scale of the clipping, respectively.
Note that the use of the proxy reward may violate our initial objective to solve the aforementioned problem because of the difference between $r_k$ and $\hat{r}_k$.
However, an appropriate setting of the offset and scale values leads to a feasible beam-tracking policy, which is validated in the numerical evaluation.

\subsection{State Expansion for Leveraging Positional Interaction}
\label{subsec:state_expansion}

\begin{figure}[t]
  \centering
  \subfigure[$t = 0$\,ms]{\includegraphics[width=.45\columnwidth]{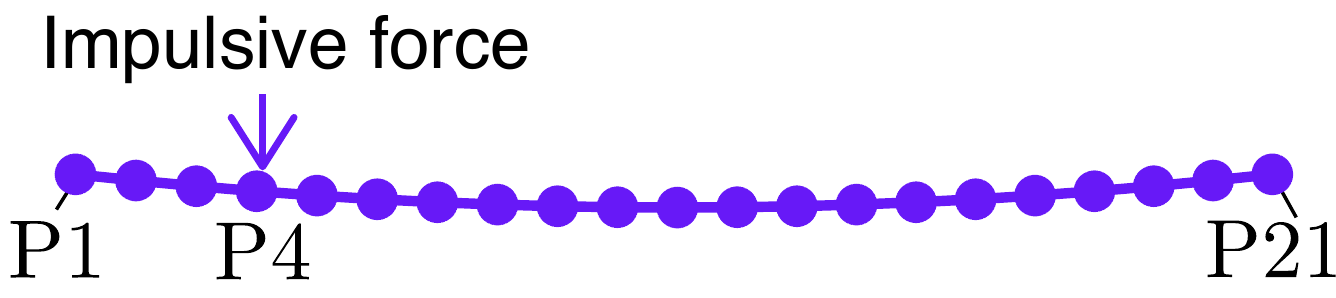}}
  \subfigure[$t = 50$\,ms]{\includegraphics[width=.45\columnwidth]{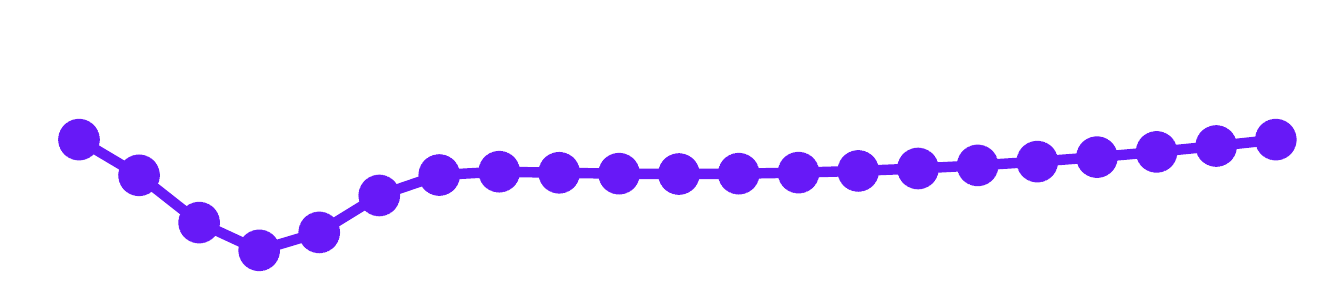}}
  \subfigure[$t = 150$\,ms]{\includegraphics[width=.45\columnwidth]{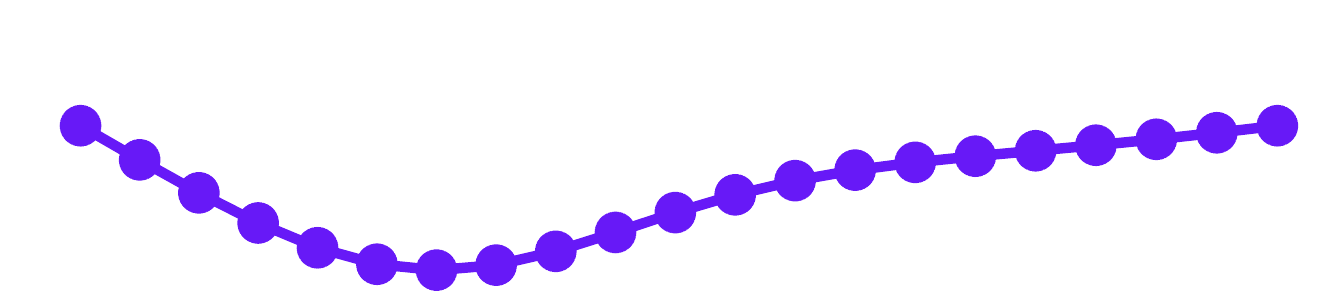}}
  \subfigure[$t = 250$\,ms]{\includegraphics[width=.45\columnwidth]{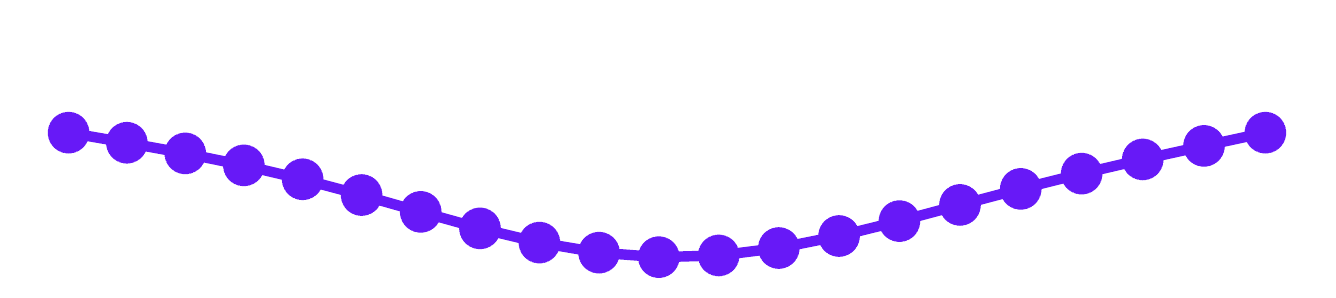}}
  \subfigure[$t = 350$\,ms]{\includegraphics[width=.45\columnwidth]{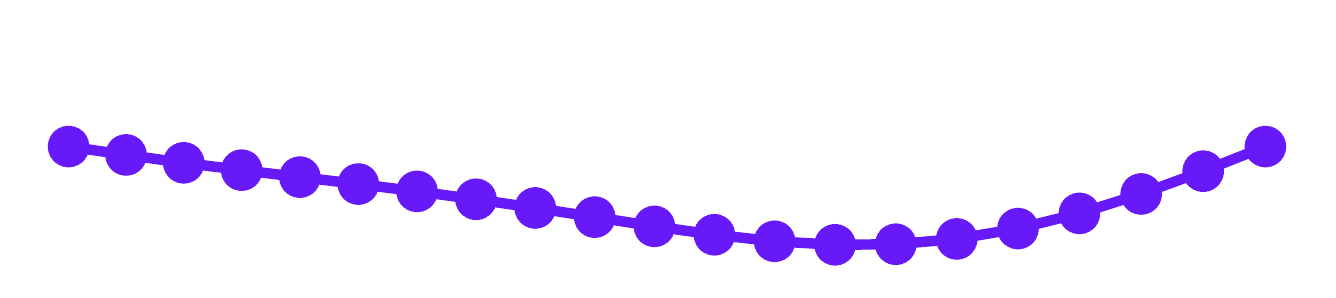}}
  \subfigure[$t = 450$\,ms]{\includegraphics[width=.45\columnwidth]{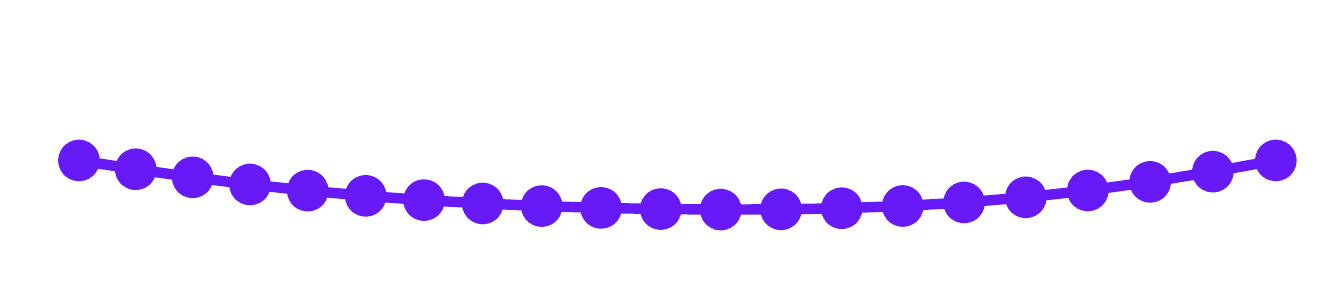}}
  \caption{Dynamics of overhead messenger wire after impulsive force with $N = 21$.}
  \label{fig:dynamics_wire}
\end{figure}

In this section, we expand the state such that the beam-forming agent leverages not only the positions and velocities of the on-wire node but also those of the points distributed on the  messenger wire.
This is motivated by the fact the points on the messenger wire interact with adjacent points via the tensile force.
Therein, the look-back positional information of the other on-wire points is also informative to predict the current positions of the on-wire node, thereby enhancing the quality of the beam steering.

To illustrate such positional interaction on messenger wires, in Fig.~\ref{fig:dynamics_wire}, we plot the 2D positions of the on-wire points after impulsive forces are injected.
This dynamics is consistent with \eqref{eq:dynamics_acceleration}, \eqref{eq:dynamics_velocity}, and \eqref{eq:dynamics_position}.
As an example, we set the parameters as $k = 1000\,\mathrm{Nm^{-1}}$, $m = 10\,\mathrm{kg}$, and $N = 21$.
At $t = 0$, an impulsive force is exerted at $\mathrm{P}4$ with the force of 470\,N, i.e., for $i = 4$, $\bm{F}_i(0) = [0, 0, 470]^{\mathrm{T}}$, and $\bm{F}_i(0) = [0, 0, 0]^{\mathrm{T}}$, otherwise.
As shown in Fig.~\ref{fig:dynamics_wire}, the positions of the points on the messenger wire correlate with the past positions of adjacent points, wherein the points are displaced in accordance with the adjacent ones and form a wave as a whole.
Hence, by looking back not only at the positions/velocities of the on-wire node, but also those of the other points, the beam-forming agent predicts the current positions of the on-wire node more accurately, thereby achieving beam steering policies nearer to the optimal solution of the aforementioned problem.

\vspace{.3em}\noindent\textbf{Expanded State.}\quad
Hence, the expanded state includes the positions and velocities of the points distributed on the messenger wire.
Specifically, the expanded state at each time instant $s_{\mathrm{ex}, k}$ is given as follows.
\begin{itemize}
  \item look-back positions \textit{of multiple points on the messenger wire}, i.e., $(\bm{x}_i(k\tau - T))_{i\in\mathcal{N}_{\mathrm{sense}}}$, where $\mathcal{N}_{\mathrm{sense}}\subseteq\{1, 2, \dots, N\}$.
  \item look-back velocities of the same points on the messenger wire, i.e., $(\bm{v}_i(k\tau - T))_{i\in\mathcal{N}_{\mathrm{sense}}}$.
  \item The current main lobe steering direction $\bm b^{(k)}$, which is given by \eqref{eq:beam_direction}.
\end{itemize}

\begin{figure}[t]
  \centering
  \includegraphics[width=\columnwidth]{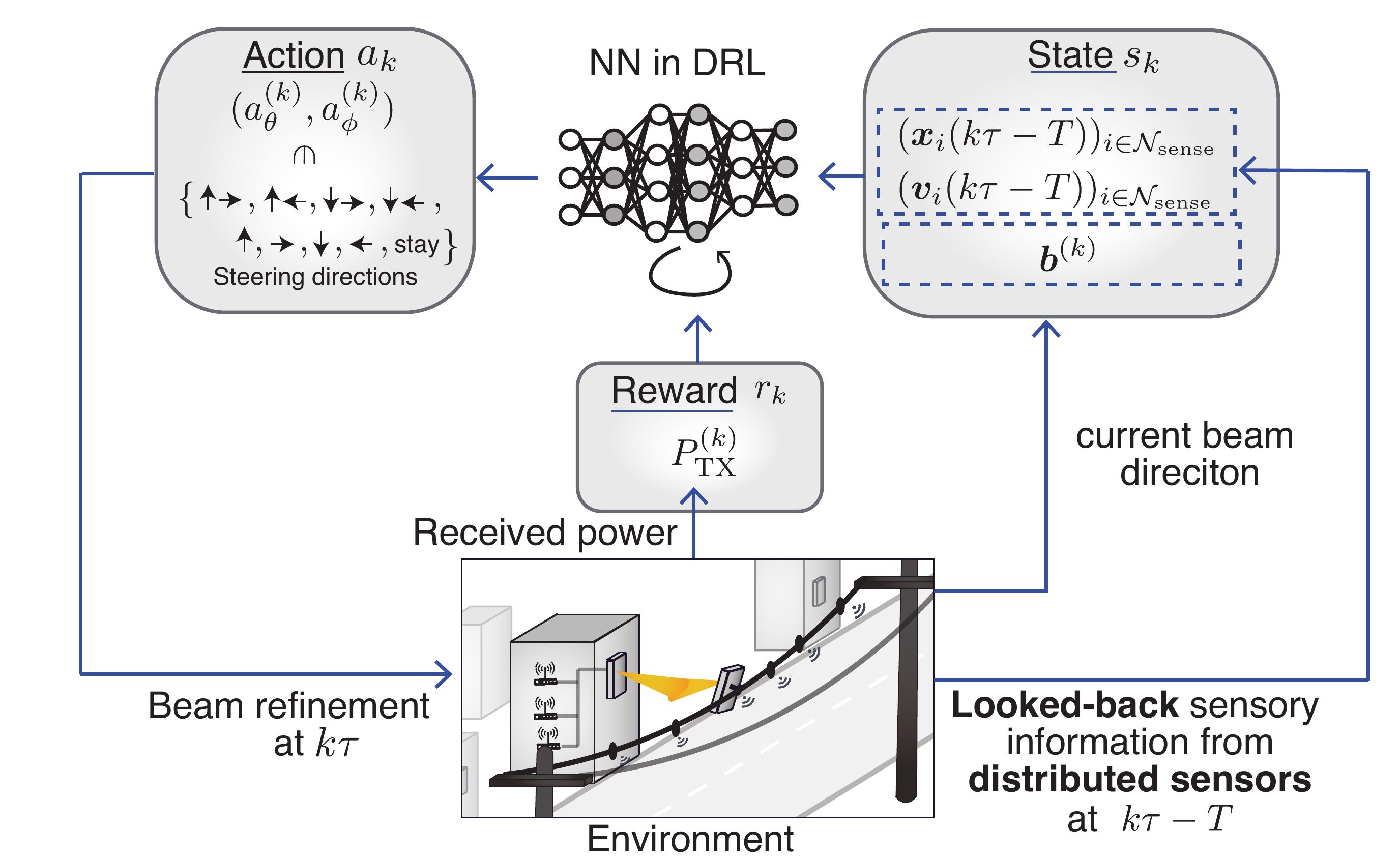}
  \caption{
    Implemented DRL-based beam tracking.
    At each time instant $k\tau$, the DRL-agent looks back sensory information from the on-wire sensors with look-back time $T$ and determines the beam steering directions while predicting the displacement of the on-wire node within $T$ by leveraging the NN.
  }
  \label{fig:RL_overview}
\end{figure}

A schematic overview of the DRL-based beam steering discussed above is presented in Fig.~\ref{fig:RL_overview}.
Note that, because our focus is on the state, action, and reward design to solve the aforementioned problem, we do not detail the specific algorithm to train the NN shown in Fig.~\ref{fig:RL_overview} and apply the algorithm detailed in the previous study \cite{mnih2015human}.
Hence, for the particulars of the algorithm, readers are encouraged to refer to the paper by Mnih \textit{et al.} \cite{mnih2015human}.

\section{Numerical Evaluation}
\label{sec:simulation}

\subsection{Setup}

\vspace{.3em}\noindent\textbf{Beam-Tracking Environment Setting.}\quad
The parameters for identifying the beam-tracking environment are summarized in Table~\ref{tbl:SimurationParam}.
The endpoints of the messenger wire and the fixed node are placed at the same height of 5\,m.
The messenger wire comprises 21 points, and the on-wire node is installed at the center of the wire, i.e., $i_{\mathrm{TX}} = 10$.
The wind speed $\bm v_\mathrm o(t)$ is given by
\begin{align}
  \label{eq:wind_speed}
  \bm v_\mathrm o(t) = \left[5\sin\frac{2\pi t}{4}, 5\sin\frac{2\pi t}{6}, 5\sin\frac{2\pi t}{8}\right]^{\mathrm{T}},
\end{align}
which follows the prior version of this work\cite{shinzaki2020deep}.

An impulsive force to the wire is occasionally applied at $\mathrm{P}4$ with the force of 470\,N, i.e., for $i = 4$, $\bm{F}_i(t_{\mathrm{force}}) = [0, 0, 470]^{\mathrm{T}}$, and $\bm{F}_i(t) = [0, 0, 0]^{\mathrm{T}}$, otherwise, where $t_{\mathrm{force}}$ is the time at which the force is applied.
When training the NN, we repeat the simulation of the on-wire dynamics for $t\in[0, 3]$, and within the duration, the impulsive force is applied once at the time determined uniformly and randomly from 1\,s, 2\,s, and 3\,s.


\begin{table}[t]
  \centering
  \caption{Parameters for Beam-Tracking Environment}
  \begin{tabular}{cc}
    \toprule
    Distance between endpoints $d_w$  & \multirow{2}{*}{$10\,\mathrm{m}$}\\
    (i.e., Distance between poles in Fig.~\ref{fig:system_model})               &  \\
    Distance between wire and building $d_r$ & \multirow{2}{*}{$5\,\mathrm{m}$}\\
     (i.e., Road width in Fig.~\ref{fig:system_model})      &                   \\
    Transmit power $P_{\mathrm{TX}}$                & $23\,\mathrm{dBm}$               \\
    Wavelength of radio waves $\lambda$                 & $5\,\mathrm{mm}$                 \\
    Receive antenna gain $G_{\mathrm{RX}}$          & $8\,\mathrm{dBi}$                \\
    Gravitational acceleration $\bm{g}$             & $[0, 0, -9.8]\,\mathrm{ms^{-2}}$ \\
    Drag constant $c_0$                             & $1\,\mathrm{s}^{-1}$             \\
    Number of points $N$                            & 21                               \\
    Mass of wire $m$                                & $10\,\mathrm{kg}$                \\
    Covariance matrix of the wind speed $\bm{V}_0$  & $0.1\bm{I}$                      \\
    Number of vertical elements $n_{\mathrm{v}}$    & 32                               \\
    Number of horizontal elements $n_{\mathrm{h}}$  & 8                                \\
    Correlation coefficient $\rho$                  & 1                                \\
    Vertical spacing distance $\Delta_{\mathrm{V}}$            & $2.5\,\mathrm{mm}$               \\
    Horizontal spacing distance $\Delta_{\mathrm{H}}$          & $2.5\,\mathrm{mm}$               \\
    Point installed on-wire node                    & $\mathrm{P}10$                   \\
    Interval between successive time instant $\tau$ & $10\,\mathrm{ms}$                \\
    Refinement angles $A$  & $1^\circ$             \\
    \bottomrule
  \end{tabular}
  \label{tbl:SimurationParam}
\end{table}

\vspace{.3em}\noindent\textbf{Beam-Tracking Agent Setting.}\quad
We detail the parameter setting for the beam-tracking agent.
Following the DRL algorithm \cite{mnih2015human}, the beam-tracking agent trains an NN that outputs an estimator of the optimal action-value $Q^{\star}(s, a)$, which is defined by the expectation of the cumulative sum of the discounted reward following the optimal policy, i.e.,
\begin{align}
  \label{eq:action_value}
  Q^{\star}(s, a)\coloneqq\max_{\pi}\mathbb{E}_{\pi}\left[\sum_{k' = 0} ^{\infty} \gamma^{k'} r_{k + k' + 1}\,\middle|\,s_k = s, a_k = a\right],
\end{align}
where $\gamma$ is the discount factor, which is assigned a value of 0.99.
The term $\pi:\mathcal{S}\to\mathcal{A}$ denotes the arbitral policy, where $\mathcal{S}$ and $\mathcal{A}$ are the state and action sets, respectively.
Note that the optimal policy $\pi^{\star}$ is determined by taking an action that maximizes the optimal action-value, i.e., by taking an action $\argmax_{a\in\mathcal{A}} Q^{\star}(s, a)$.
Hence, to obtain the optimal actions given the state information, it is sufficient to learn the optimal action-value as accurately as possible.
Let the NN that estimates the optimal action-value be denoted by $Q(s, a; \bm \theta)$ in the sequel.
The NN model $Q(s, a; \bm \theta)$ in this evaluation is multi-layer-perceptron that accepts the input of state information and outputs the action-values corresponding each action.
The network consists of three hidden layers having $128$ units.
The activation function of the hidden layers is the rectified linear unit (ReLU) $R(x)$ given by $R(x) \coloneqq \max \{x, 0\}$.
This network architecture is just a preliminary example, and however is shown to be sufficient to achieve our objective, i.e., feasibly learning an appropriate beam-tracking policies in the on-wire dynamics.

In the DRL algorithm, the NN is trained such that the difference between $Q(s_k, a_k; \bm \theta)$ and $r_k + \gamma\max_{a'} Q(s_{k + 1}, a'; \bm \theta)$ is minimized.
To quantify the difference, we leverage the following Huber Loss\cite{varga2018deeprn} as an example:
\begin{align}
  L(\bm \theta) =
  \begin{cases}
    x^2/2,     & x \leq 1;         \\
    |x| - 0.5, & \text{otherwise},
  \end{cases}
\end{align}
where $x \coloneqq Q(s_k, a_k; \bm \theta) - \bigl(r_k + \gamma\max_{a'} Q\bigl(s_{k + 1}, a_{k + 1}; \hat{\bm \theta}\bigr)\bigr)$.
Therein, $Q\bigl(\cdot, \cdot; \hat{\bm \theta}\bigr)$ is the proxy NN, which is known as the ``target network'' and is updated periodically to stabilize the training\cite{mnih2015human}.
To minimize the above loss function, we utilize the well-known Adam algorithm\cite{sutskever2013importance} with a learning rate of $10^{-4}$ per 300 time steps.
This parameter update with the Adam optimizer is performed over the 2048 past experiences of $(s_k, a_k, r_k, s_{k + 1})$ randomly sampled from a replay buffer with the mini-batch size of 32 and the number of epochs of 8, which is iterated by four times while varying the sampled experiences.
The target network is updated each 3000 time steps.
The process of training the network required 100,000 time steps in total, and the policy determined via the trained NN was tested while running 300 time steps after every update of the NN.

During training, DRL requires the exploration of unexplored actions.
To do this, we employ an $\epsilon$-greedy policy that select the action that maximizes $Q(s, a; \bm \theta)$ with the probability of $1 - \epsilon$, selecting actions uniformly randomly otherwise.
In this experiment, $\epsilon$ is set as 0.2 as an example.
Meanwhile, while testing the learned policy, we set $\epsilon$ as 0, which corresponds to select an action in a greedy fashion according to the learned optimal action values.

\subsection{Beam-Tracking Performance without Knowing Positional Interaction}
We evaluate the learned beam-tracking policy without knowing the positional interaction on the wire, i.e., without the state expansion discussed in Section~\ref{subsec:state_expansion}.
These results showed that a longer look-back time $T$ results in frequent beam misalignment particularly in the case of disturbances caused by impulsive forces to the wire, which seamlessly emphasizes the importance of the aforementioned state expansion in leveraging the positional interaction.

\begin{figure}[t]
  \centering
  \subfigure[Instantaneous received power.]{\includegraphics[width=0.7\columnwidth]{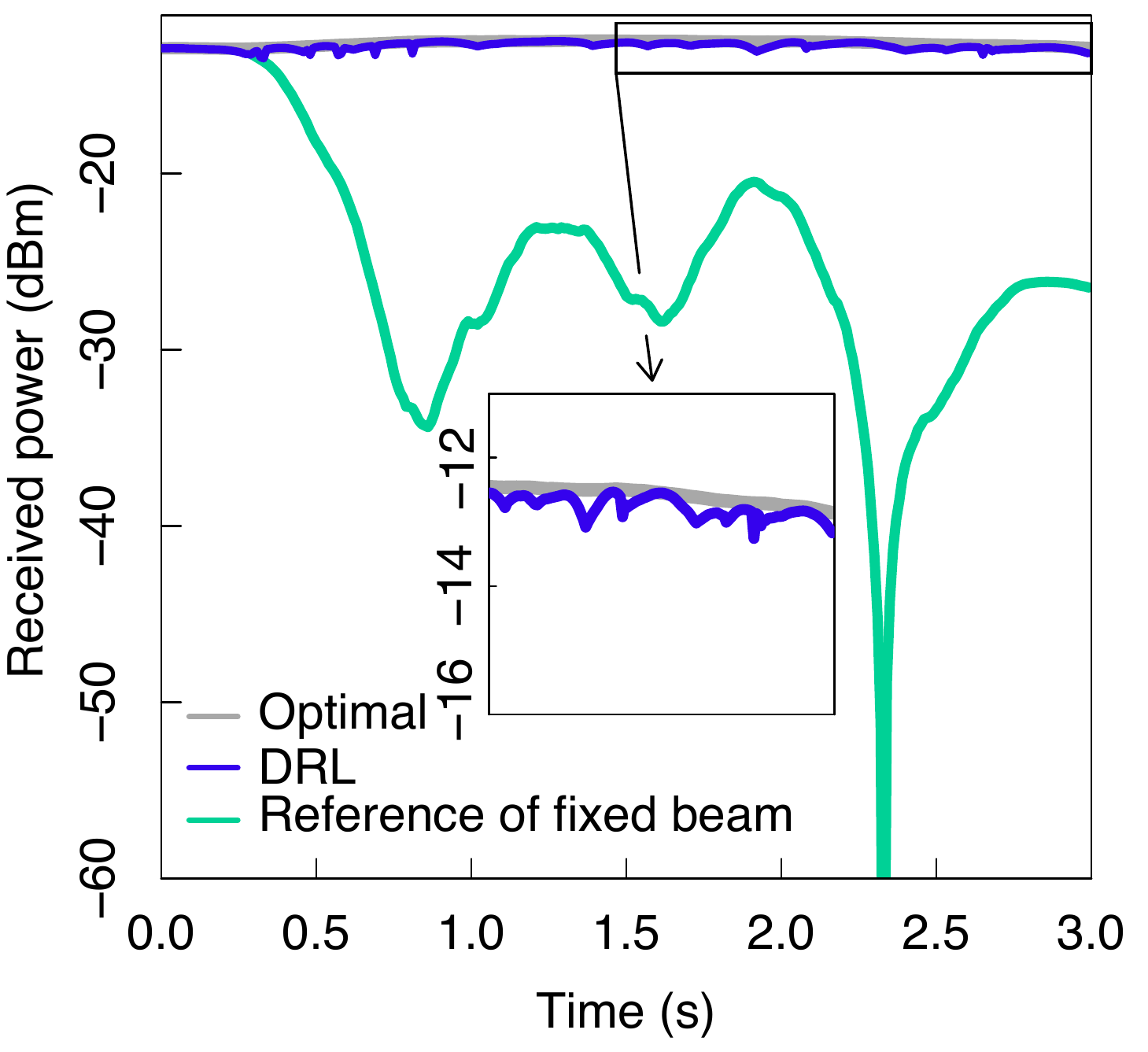}}\vspace{.5em}
  \subfigure[Difference from optimal angle.]{\includegraphics[width=0.7\columnwidth]{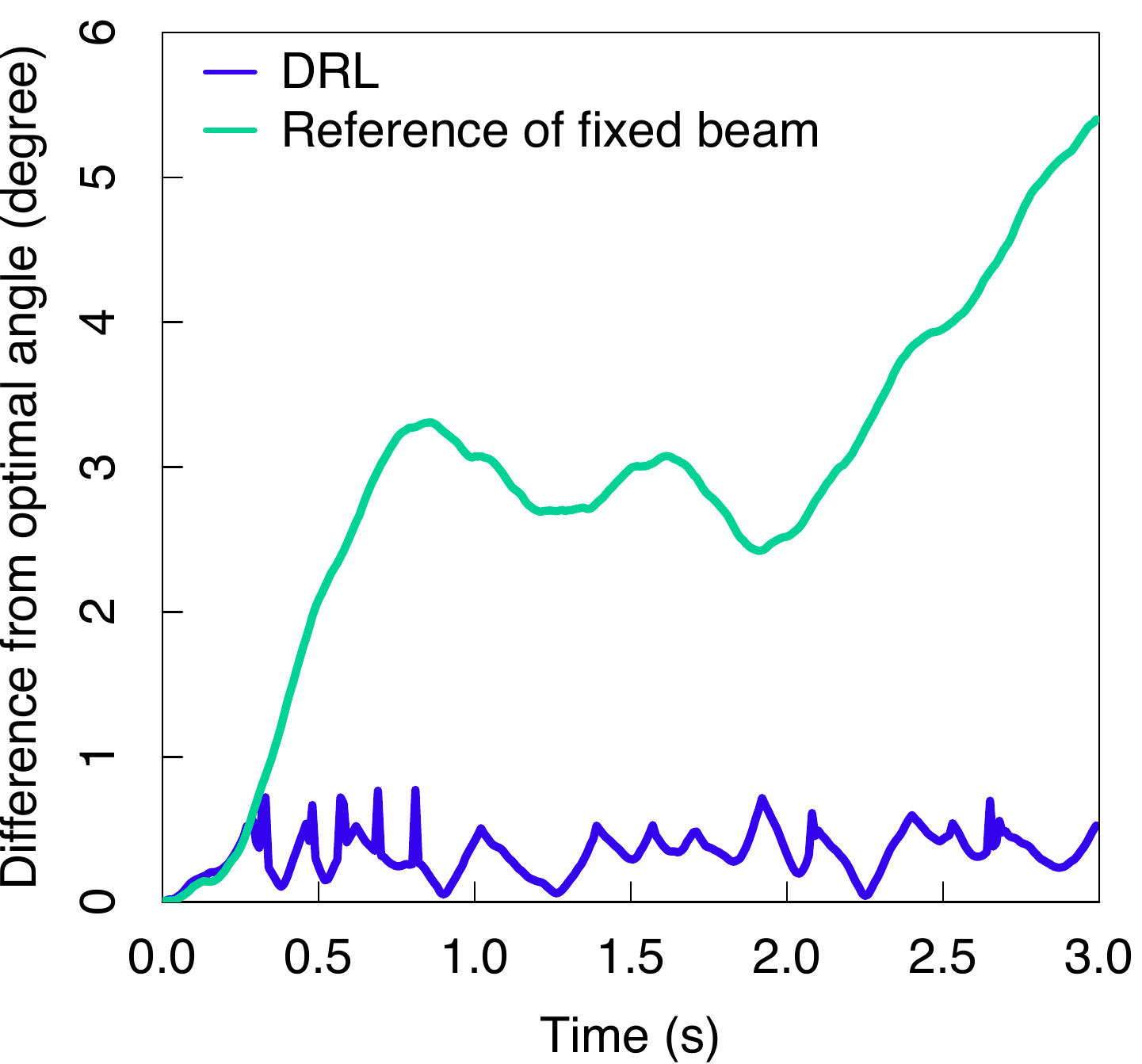}}
  \caption{Time-varying received power and difference from optimal policy in terms of the steering angle when the position of the on-wire node is affected only by wind perturbation ($m = 10\,\mathrm{kg}$, $k_0 = 1000\,\mathrm{Nm^{-1}}$, and $T = 20\,\mathrm{ms}$).}
  \label{fig:quick_analysis}
\end{figure}

\vspace{.3em}\noindent\textbf{Robustness Against Wind Perturbations.}\quad
First, we confirm that our DRL-based beam tracking learns an appropriate beam-tracking policy for a simple case in which the on-wire node is perturbed only by the wind.
Fig.~\ref{fig:quick_analysis} shows the instantaneous angle difference from the optimal angles and received power in an exemplary situation where $m = 10\,\mathrm{kg}$, $k_0 = 1000\,\mathrm{Nm^{-1}}$, and $T = 20\,\mathrm{ms}$.
As a reference, we plot the results from the beam fixed at an initial angle to allow ones to capture how seriously wind perturbations affect the received power.
As shown in Fig.~\ref{fig:quick_analysis}(a), with the fixed beam, the received power is drastically degraded, which indicates the importance of beam tracking in the on-wire mmWave node.
Meanwhile, the received power in the DRL-based beam tracking comes close to the optimal values, and the difference between them is within 2\,dB.
This indicates that the angle of departure relative to the main lobe steering angle is within half-power beam width, and in this sense, the beam is appropriately steered against the wind perturbations.
Note that the near-optimal solution can also be achieved by conventional exhaustive search for short periods in which possible beam directions are rapidly scanned to find the beam steering direction with maximum received power.
Nevertheless, the importance of this result lies in the fact that \textbf{an appropriate beam-tracking policy can be feasibly learned based on the look-back information about the position and velocity} without scanning channel information.

\begin{figure}[t]
  \centering
  \subfigure[Average received power vs. mass of overhead messenger wire $m$ ($k_0 = 1000\,\mathrm{Nm^{-1}}$).]{\includegraphics[width=0.66\columnwidth]{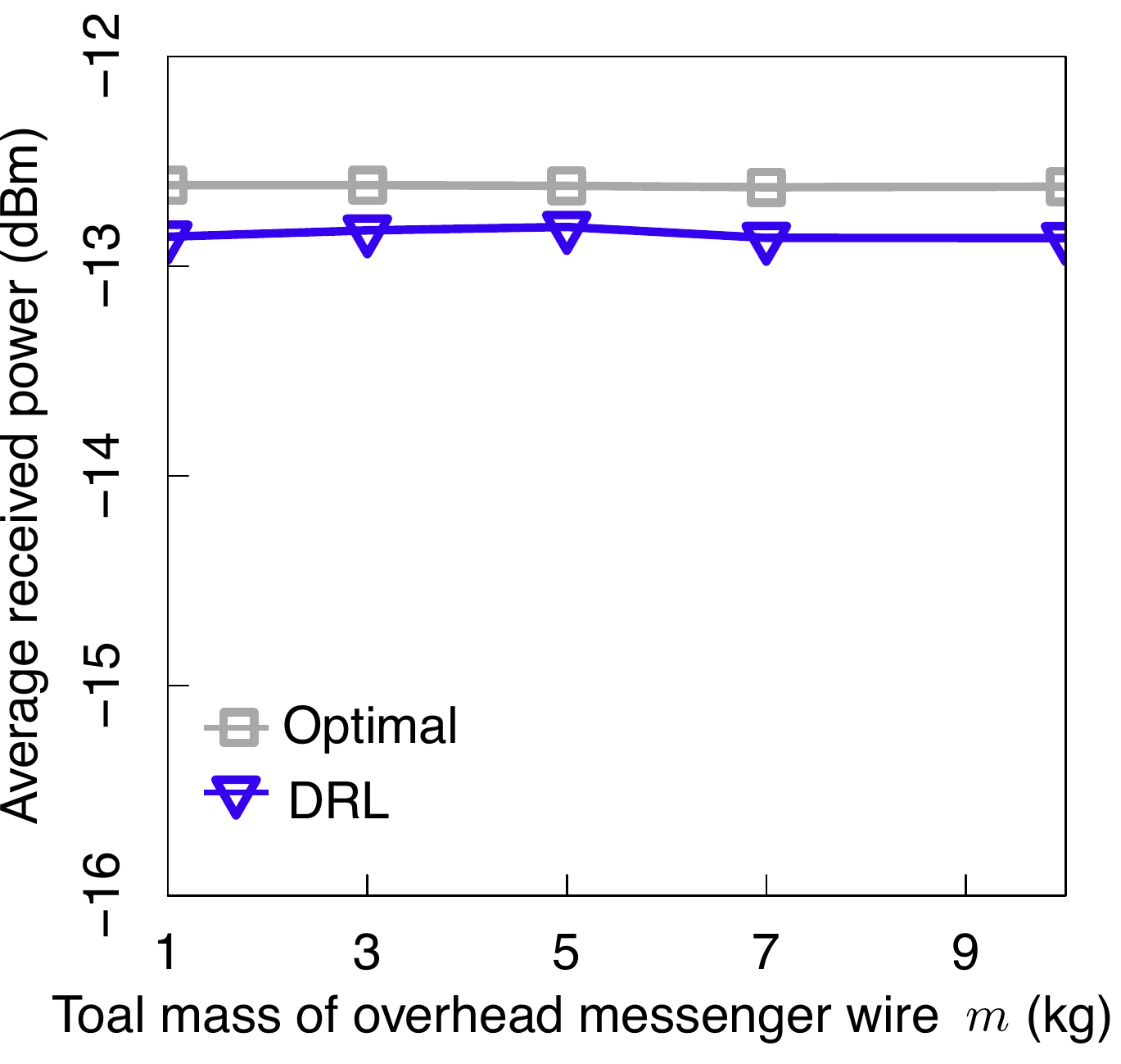}}
  \subfigure[Average received power vs. spring constant $k_0$ ($m = 10\,\mathrm{kg}$).]{\includegraphics[width=0.66\columnwidth]{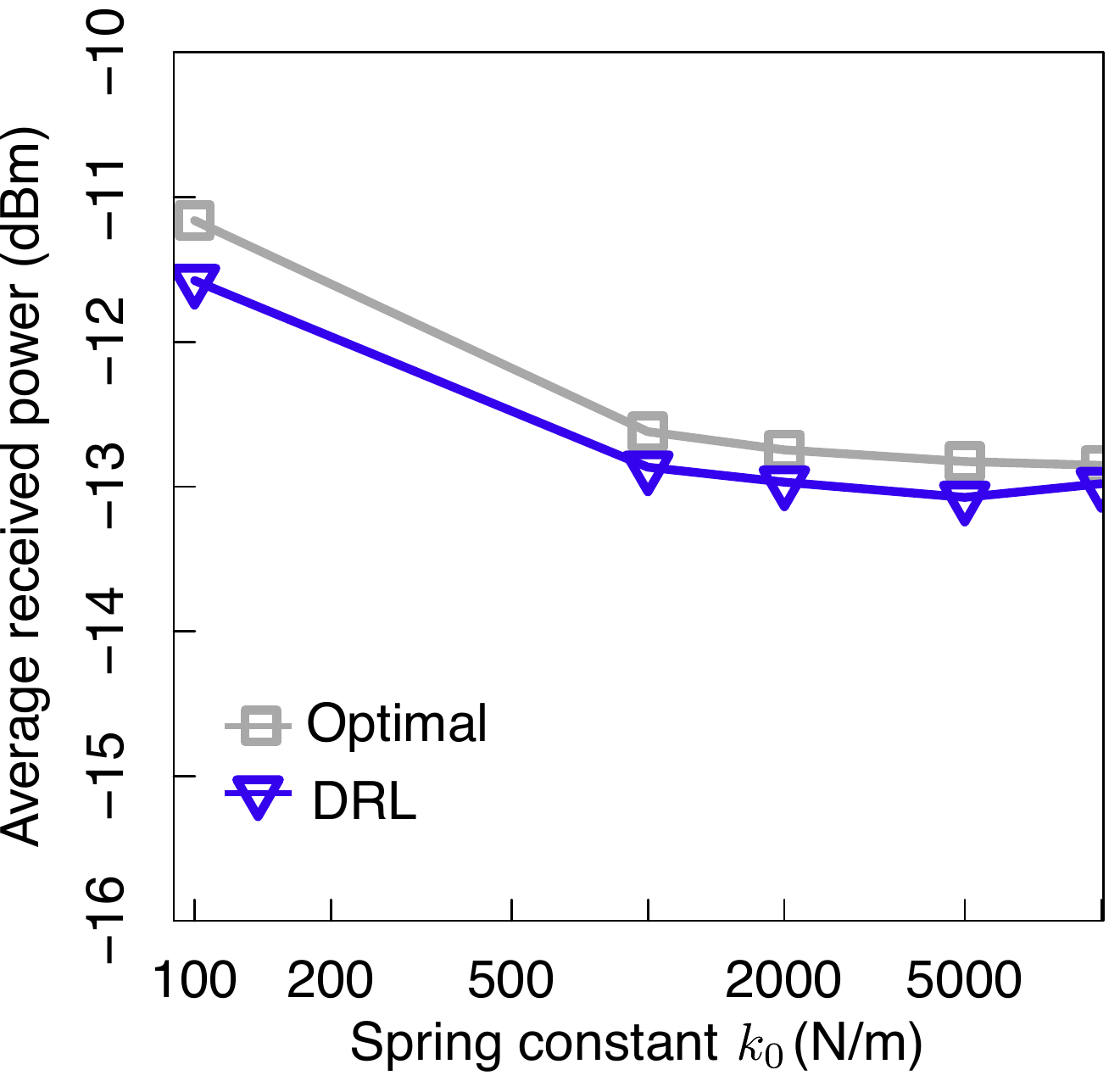}}\vspace{-0.5em}
  \caption{Dependency of wire characteristics when the position of the on-wire node is affected only by wind perturbation (Look-back time $T = 80\,\mathrm{ms}$).}
  \label{fig:wind_one_point_various_parameters}
\end{figure}

In Fig.~\ref{fig:wind_one_point_various_parameters}, we investigate the feasibility of learning appropriate beam-tracking policies for various wire characteristics.
Specifically, the dependency of the mass of the messenger wire $m$ is shown in Fig.~\ref{fig:wind_one_point_various_parameters}(a) whereas that of the spring constant $k_0$ is shown in Fig.~\ref{fig:wind_one_point_various_parameters}(b).
Both figures show that the received powers of the DRL-based beam-tracking come close to the optimal values, with the difference between them being less than 0.5\,dB, which is a much smaller difference than without beam tracking.
Hence, against wind perturbations (without impulsive forces to the wire), we can feasibly learn an appropriate beam-tracking policy for various wire characteristics.

\begin{figure}[t]
  \centering
  \subfigure[Instantaneous received power.]{\includegraphics[width=0.7\columnwidth]{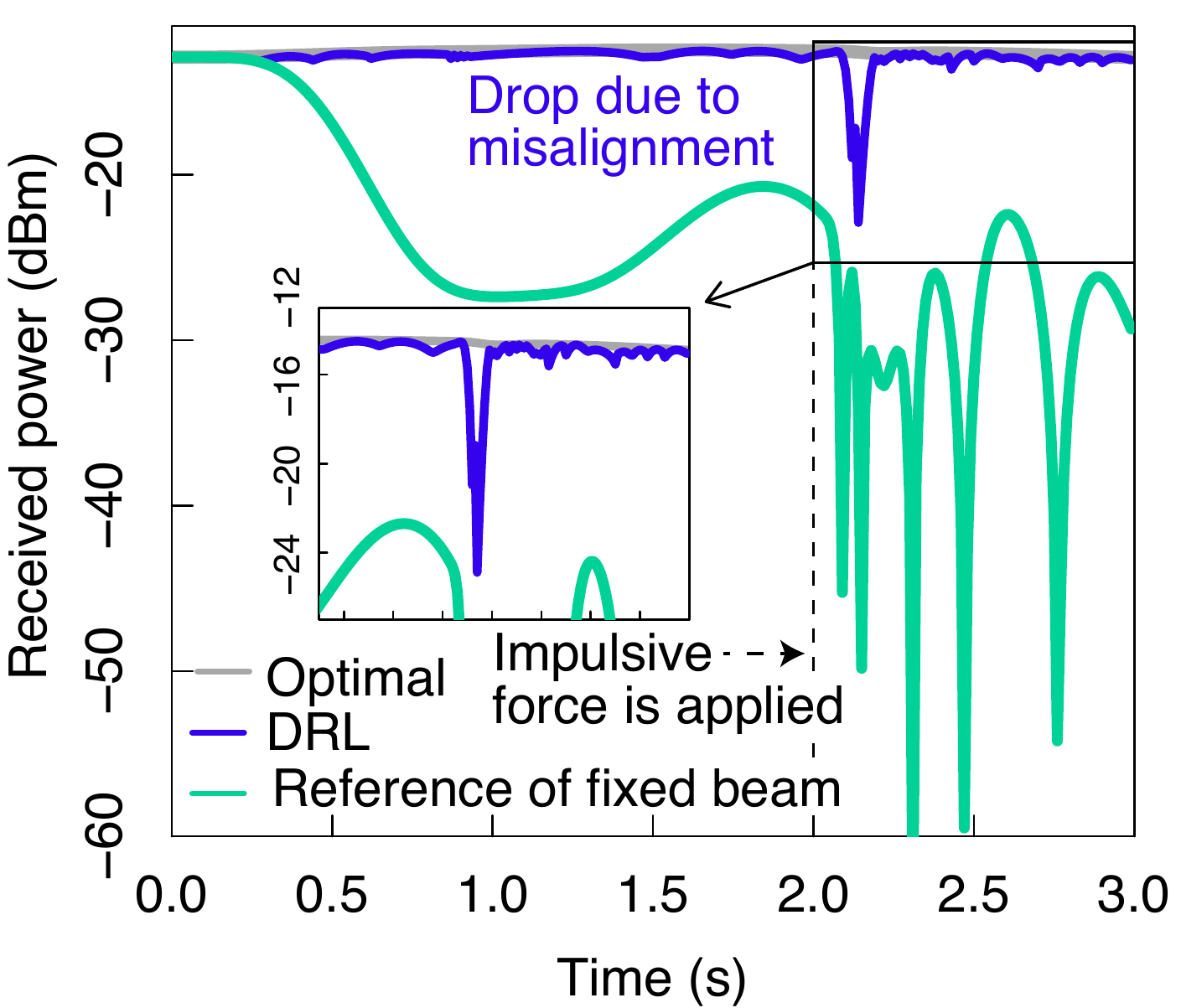}}\vspace{1em}
  \subfigure[Difference from optimal angle.]{\includegraphics[width=0.7\columnwidth]{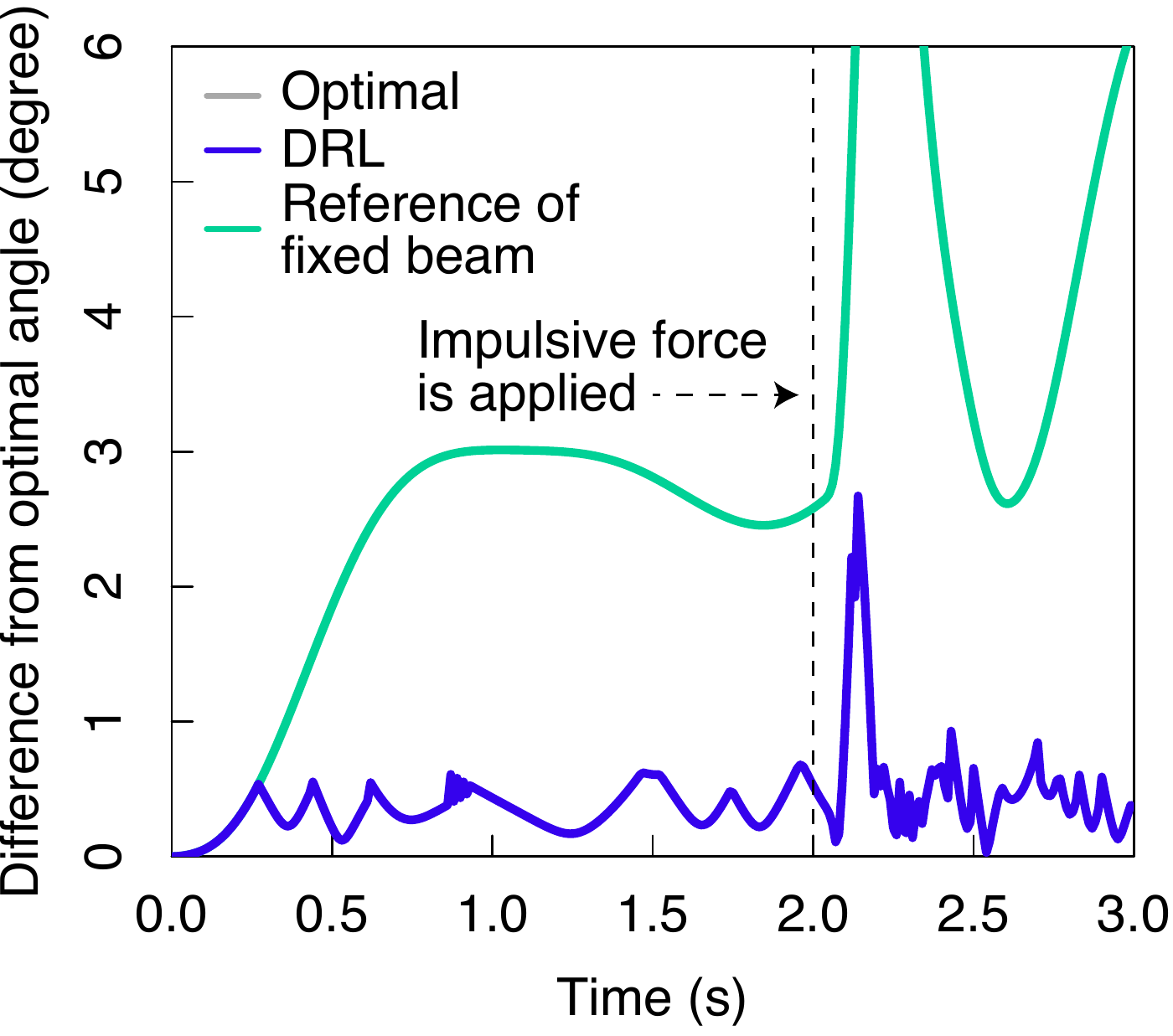}}
  \caption{Time-varying received power and difference from optimal policy in terms of steering angle in the  position of the on-wire node affected by both wind perturbations and impulsive forces to the wire ($m = 10\,\mathrm{kg}$, $k_0 = 1000\,\mathrm{Nm^{-1}}$, and $T = 80\,\mathrm{ms}$).}
  \label{fig:collsision_missalignment}
\end{figure}

\vspace{.3em}\noindent\textbf{Robustness Against Displacements From Impulsive Force.}\quad
We evaluate the performance ot the learned beam-tracking policy in a situation in which the position of the on-wire node is affected by both wind perturbations and impulsive forces.
Fig.~\ref{fig:collsision_missalignment} shows the instantaneous received power and the angle differences from the optimal angles in the same parameter setting as Fig.~\ref{fig:quick_analysis}.
In Fig.~\ref{fig:collsision_missalignment}(a), we can observe that the received power is degraded by approximately 10\,dB after the impulsive force is applied at 2\,s.
This degradation in received power is due to beam misalignment, which is confirmed in Fig.~\ref{fig:collsision_missalignment}(b), in which the difference from the optimal policy is larger in terms of the steering angle.
This beam misalignment is attributed to the fact that the look-back position of the on-wire node lacks informative features for predicting the current on-wire node position.
Indeed, the variation in the position of the on-wire node caused by the impulsive force is more rapid than that caused by wind perturbations.
During such a rapid variation, the look-back positions of the on-wire node and the current positions are decorrelated, and hence, using only the look-back positions, it would not be possible for the agent to predict the current on-wire node positions, resulting in beam misalignment.
In the next section, we demonstrate that this beam misalignment can be prevented by utilizing the aforementioned expanded state, that is, by utilizing the positions and velocities of the points distributed on the messenger wire as state information.

\begin{figure}[t]
  \centering
  \includegraphics[width=0.8\columnwidth]{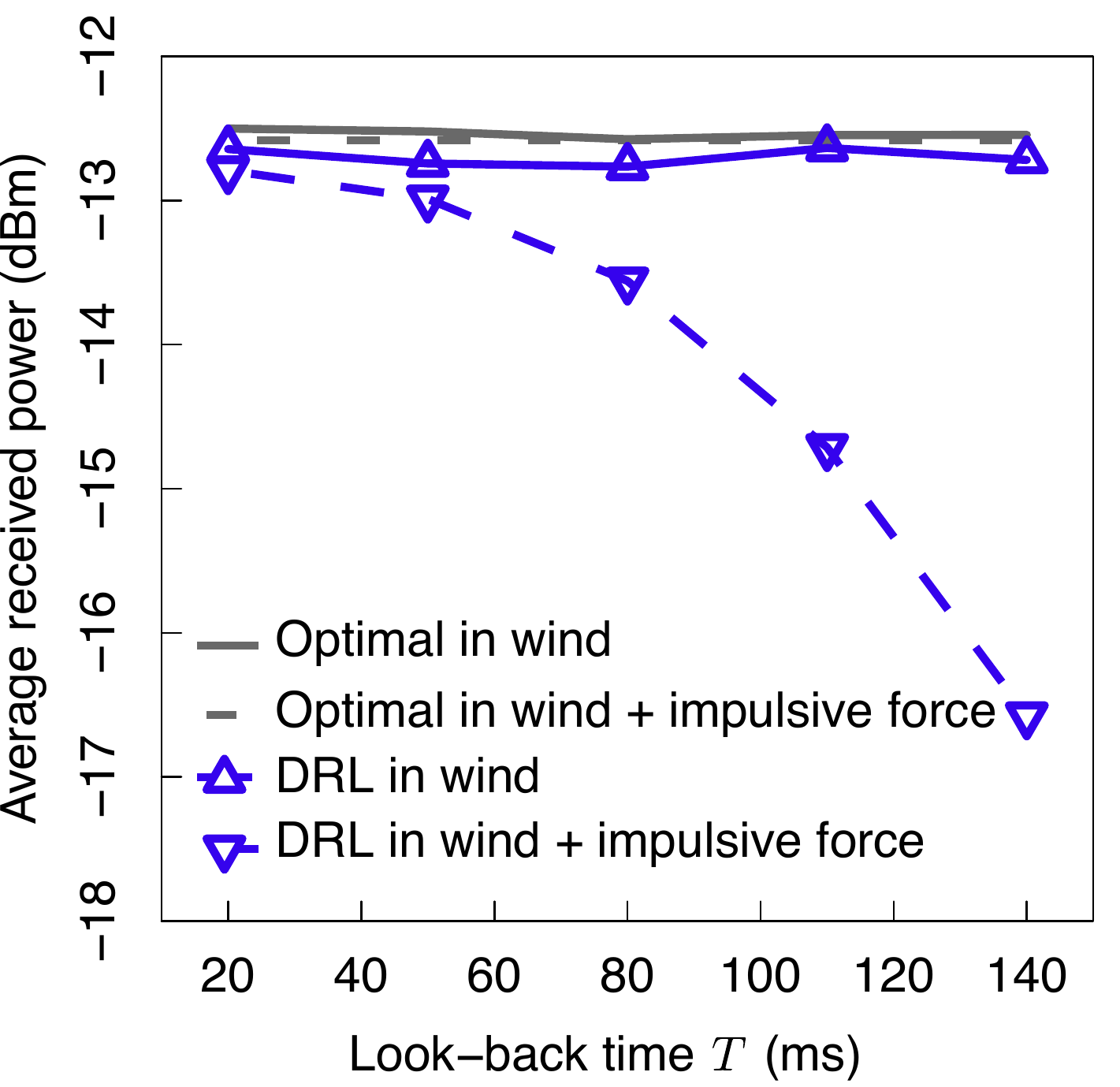}
  \caption{Received power averaged for 300\,ms after an impulsive force is applied to the wire in the use of only the position and veocity of the on-wire node.}
  \label{fig:avg_rss_delay_w_wo_bird}
\end{figure}

During the rapid displacement of the on-wire node as a result of the impulsive force, the look-back positions of the on-wire node become less informative to predict current positions, and the average received power decreases accordingly.
We confirm this in Fig.~\ref{fig:avg_rss_delay_w_wo_bird}, which shows the received power averaged for 300\,ms after the impulsive force is applied.
As a reference, we also plotted the average received power within the same time duration for a situation in which no impulsive forces are applied.
Fig.~\ref{fig:avg_rss_delay_w_wo_bird} clearly shows that the learned beam-tracking policy performs worse in terms of the average received power as the look-back time becomes longer, which can be explained by the aforementioned reasons.

\subsection{Beam-Tracking Performance with Knowing Positional Interaction}

In this section, we confirm that the aforementioned beam misalignment can be avoided by utilizing the expanded state.
This allows the beam-training agent to be informed not only of the dynamics of the on-wire node but also that of the multiple points of the messenger wire.
We consider the impulsive force as the cause of the perturbation that affects the position of the on-wire node because, under such conditions, the beam misalignment was shown to be more severe (refer the previous section).
As an example, we use the position and velocity of nine points on the messenger wire.
The indices of these points are $2, 4, 6, 8, 10, 12, 14, 16, 18$, i.e., $\mathcal{N}_{\mathrm{sense}} = \{\,2n\,|\, n\in\mathbb{N}, 1\leq n\leq 9 \,\}$.

\begin{figure}[t]
  \centering
  \subfigure[Instantaneous received power.]{\includegraphics[width=0.83\columnwidth]{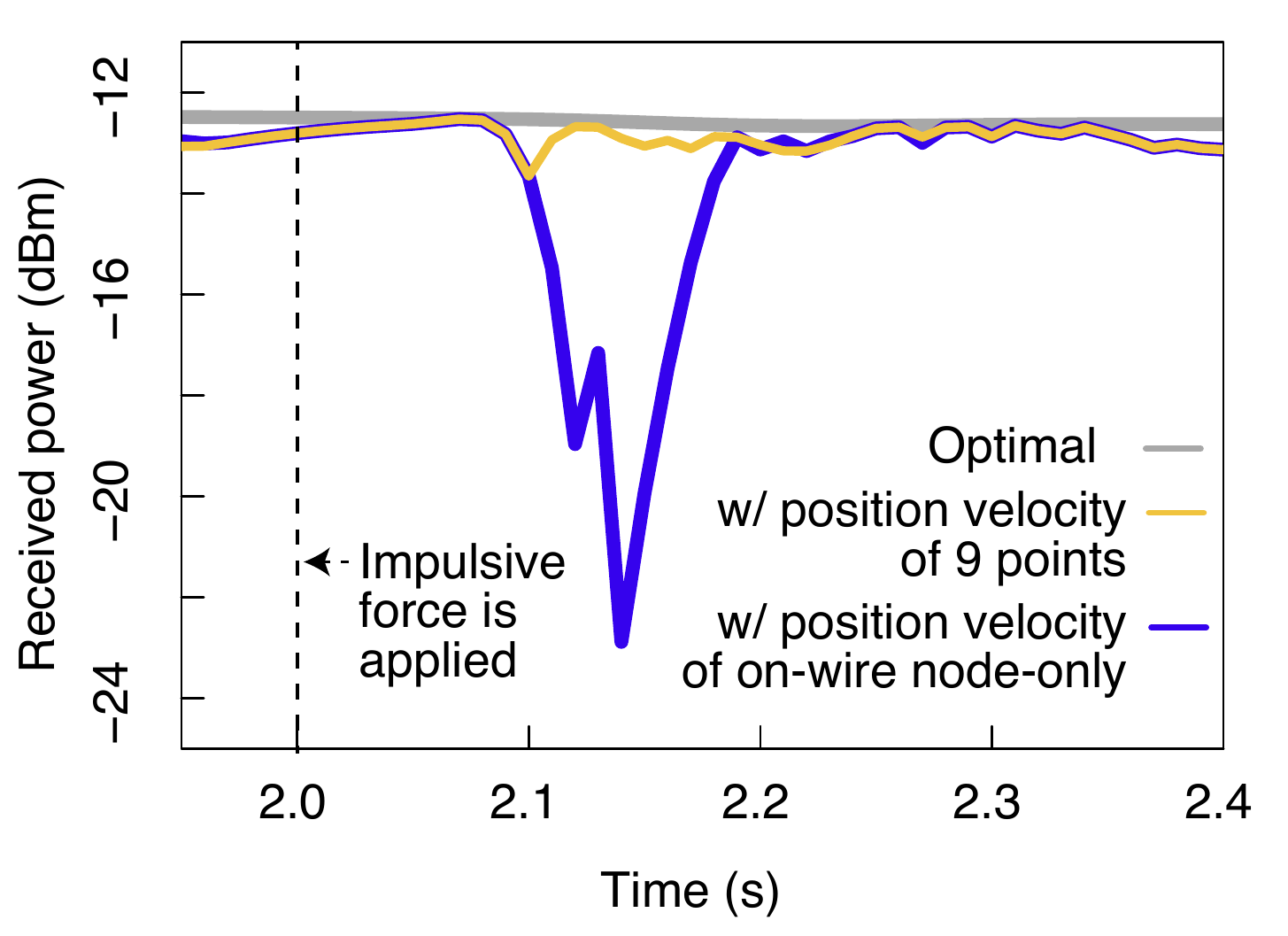}}\hspace{1em}
  \subfigure[Difference from optimal angle.]{\includegraphics[width=0.83\columnwidth]{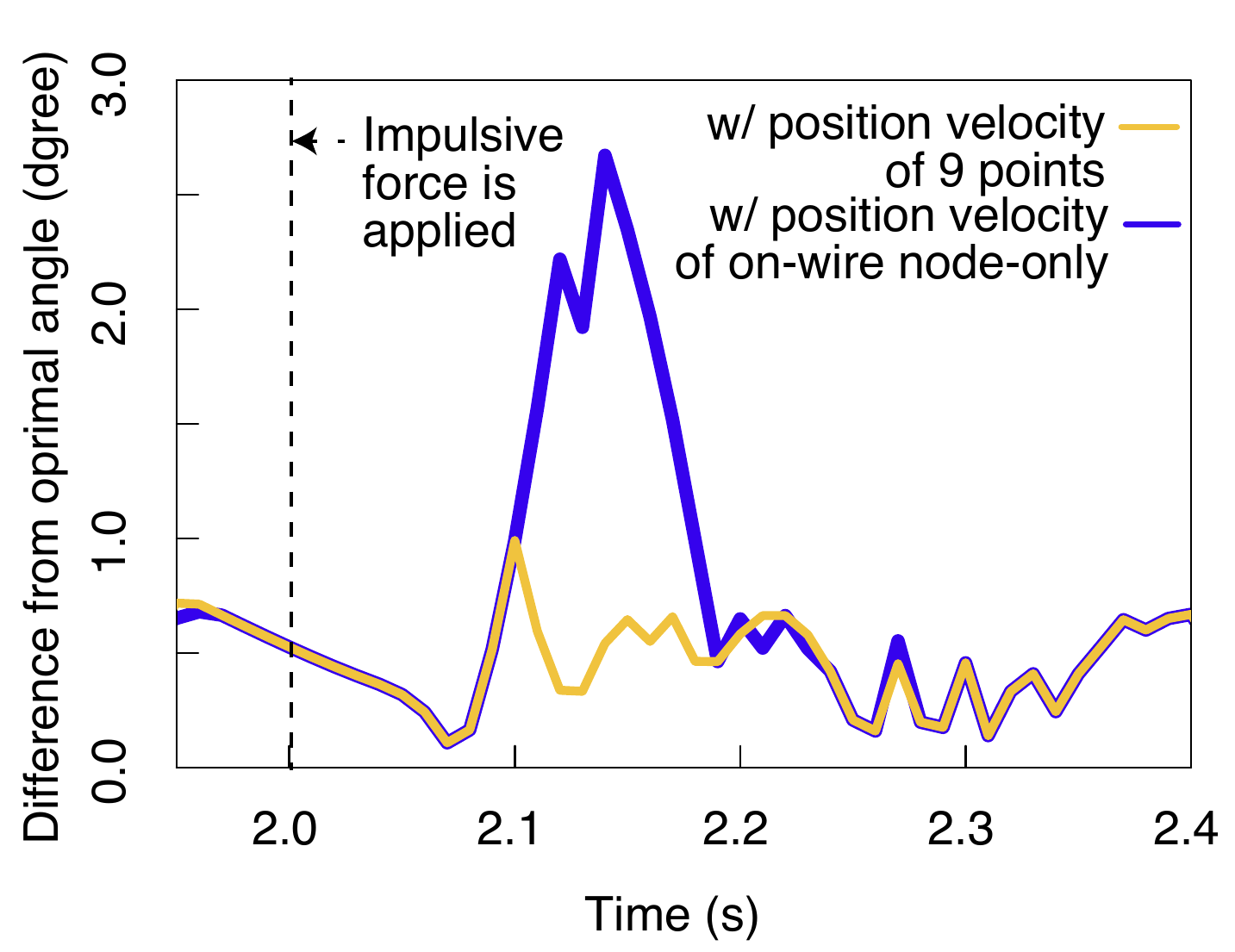}}
  \caption{Comparison of the state definition (use of only the position and velocity of the on-wire node vs. that for nine points on the messenger wire) in terms of the time-varying received power and difference from the optimal angle ($m = 10\,\mathrm{kg}$, $k_0 = 1000\,\mathrm{Nm^{-1}}$, and $T = 80\,\mathrm{ms}$).}
  \label{fig:tiemseries_rss_angle_1p_9p}
\end{figure}

The results in Fig.~\ref{fig:tiemseries_rss_angle_1p_9p} confirm that the beam misalignment in Fig.~\ref{fig:collsision_missalignment} can be feasibly avoided by leveraging the expanded state that utilizes the position and velocity of nine points on the wire.
Fig.~\ref{fig:tiemseries_rss_angle_1p_9p} compares the situation with and without the state expansion in terms of the instantaneous received power and difference from the optimal steering angle.
Note that the result in Fig.~\ref{fig:tiemseries_rss_angle_1p_9p} without the state expansion is same as that in Fig.~\ref{fig:collsision_missalignment}.
Fig.~\ref{fig:tiemseries_rss_angle_1p_9p}(a) shows that the received power for the beam tracking with the state expansion does not decrease as severely as that without the state expansion.
This is because beam misalignment is avoided, as shown in Fig.~\ref{fig:tiemseries_rss_angle_1p_9p}(b), which shows the less difference from the optimal steering angle.
These results are understood as discussed in Section~\ref{subsec:state_expansion}: the looked-back positions and velocities of multiple points on the messenger wire include an informative feature that can be used to predict sudden variation in the current position of the on-wire node as a result of the impulsive force.
Hence, the appropriate beam-tracking policy could be learnable even when the position of the on-wire node varies suddenly because of the impulsive force.

\begin{figure}[t]
  \centering
  \includegraphics[width=0.8\columnwidth]{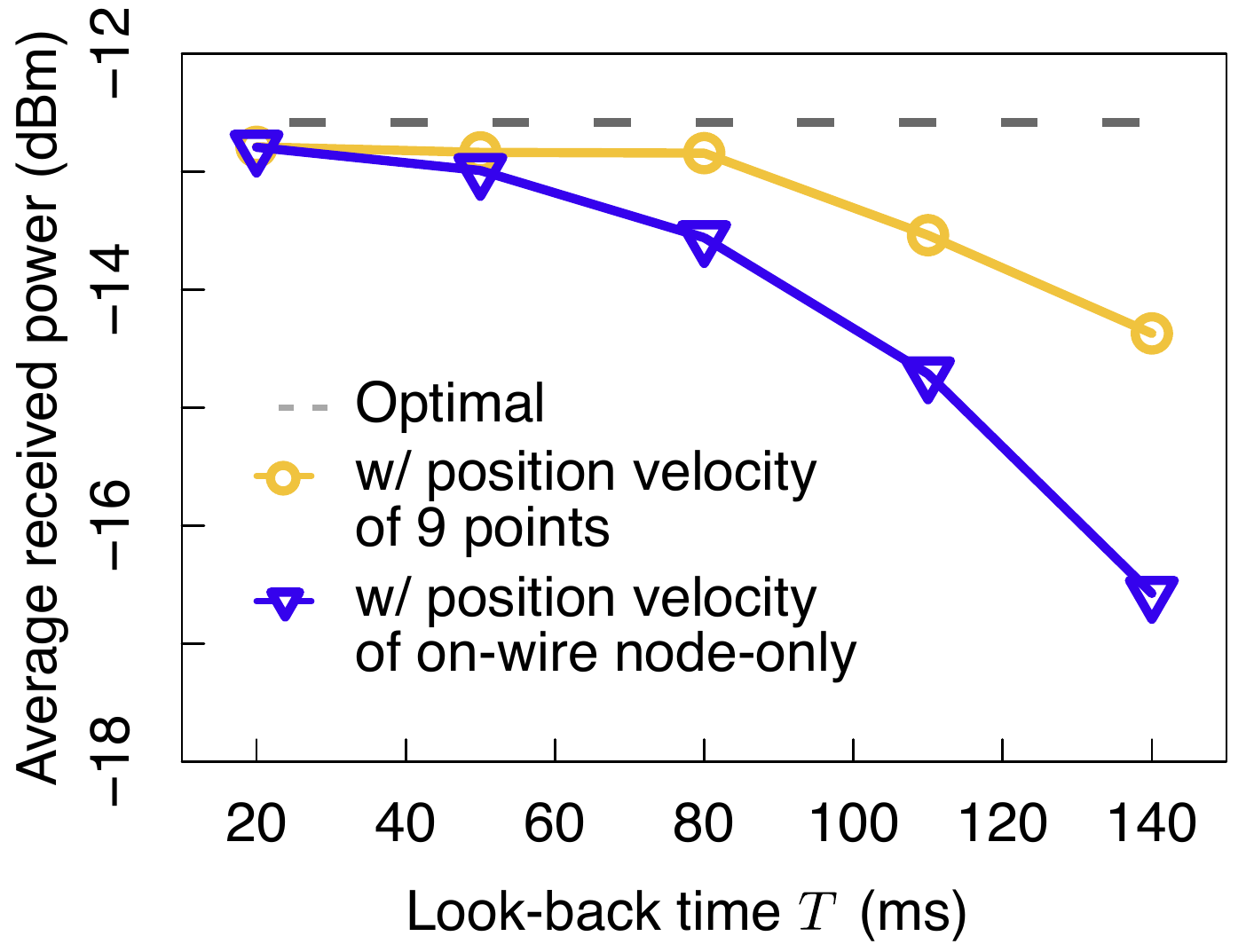}
  \caption{Comparison of the state definition (use of only the position and velocity of the on-wire node vs. that for nine points on the messenger wire) in terms of the average received power for 300\,ms after the impulsive force is applied to the wire.}
  \label{fig:avg_rss_vs_delay_1p_9p}
\end{figure}

We also confirm that the degradation of the average received power because of the increase in the look-back time $T$ can be mitigated by using the expanded state.
In Fig.~\ref{fig:avg_rss_vs_delay_1p_9p}, we show the received power averaged for 300\,ms after the impulsive force is applied.
Again, the results without state expansion shown in Fig.~\ref{fig:avg_rss_vs_delay_1p_9p} are the same as those in Fig.~\ref{fig:avg_rss_delay_w_wo_bird}.
The results in Fig.~\ref{fig:avg_rss_vs_delay_1p_9p} show that the average received power for the beam tracking with the state expansion is higher than that without.
This is attributable to the aforementioned avoidance of the beam-misalignment achieved by knowing the positional interaction on the wire.

\section{Conclusion}
\label{sec:conclusion}
In this paper, we investigated the feasibility of tracking mmWave directional beams under different types of perturbations specific to on-wire placement based on historical information about the position and velocity.
This beam-tracking problem presents a challenge owing to the complicated dynamics of messenger wires subject to a tensile force, wind perturbations, and impulsive forces applied to the wire.
We implemented a DRL-based beam-tracking framework to learn the relationship between these position and velocity values of a mmWave node and an appropriate beam steering angle.

Numerical evaluations to reproduce the wire-like dynamics demonstrated the feasibility of the DRL-based beam-tracking framework.
It was shown that, against wind perturbations, beam misalignment is feasibly avoided based on the historical positions and velocities of the mmWave node.
Moreover, it was shown that beam misalignment resulting from an impulsive force to the wire cannot necessarily be avoided by using only historical positions and velocities; however, it can be feasibly avoided by leveraging the positions and velocities of several points on the messenger wire.

It should finally be noted that this study was conducted with the objective of allowing ones to make the most of mmWave radios including 5G NR by exploring a new installation place potentially beneficial to ensure LOS connectivities.
This sheds light on gaining the flexibility in the physical node deployment, and in this sense, we believe that the insight provided by this study contributes to the future on-going network deployment of the 5G and beyond.


\bibliographystyle{IEEEtran}
\bibliography{IEEEabrv,main}

\begin{IEEEbiography}[{\includegraphics[width=1in,height=1.25in,clip,keepaspectratio]{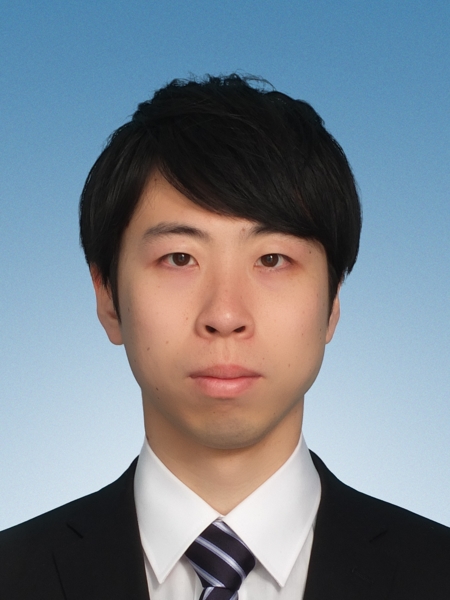}}]{Yusuke~Koda}
  received the B.E.\@ degree in electrical and electronic engineering from Kyoto University in 2016
  and the M.E.\@ degree at the Graduate School of Informatics from Kyoto University in 2018.
  In 2019, he visited Centre for Wireless Communications, University of Oulu, Finland to conduct collaborative research.
  He is currently studying toward the Ph.D.\@ degree at the Graduate School of Informatics from Kyoto University.
  He received the VTS Japan Young Researcher's Encouragement Award in 2017 and TELECOM System Technology Award in 2020.
  He was a Recipient of the Nokia Foundation Centennial Scholarship in 2019.
  He is a member of the ACM and IEICE.
\end{IEEEbiography}
\begin{IEEEbiography}[{\includegraphics[width=1in,height=1.25in,clip,keepaspectratio]{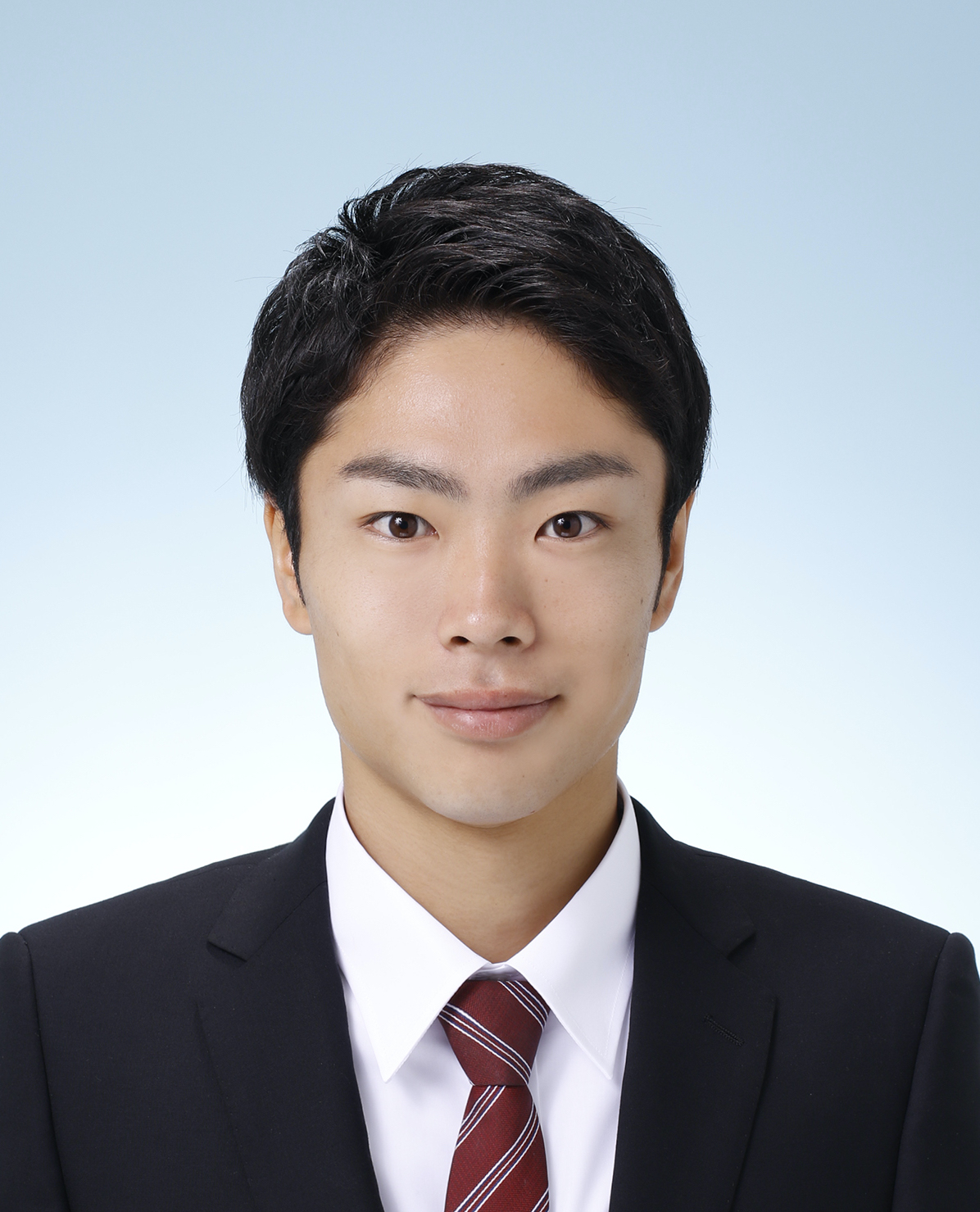}}]{Masao Shinzaki}
  received the B.E. degree in electrical and electronic engineering from Kyoto University in 2019.
  He is currently studying toward the M.E. degree at the Graduate School of Informatics from Kyoto University.
  He is a member of the IEICE.
\end{IEEEbiography}

\begin{IEEEbiography}[{\includegraphics[width=1in,height=1.25in,clip,keepaspectratio]{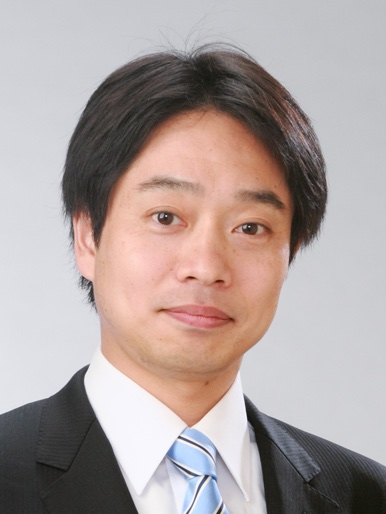}}]{Koji Yamamoto}
  (S'03--M'06--SM'20) received the B.E. degree in electrical and electronic engineering from Kyoto University in 2002, and the master and Ph.D.\ degrees in Informatics from Kyoto University in 2004 and 2005, respectively.
  From 2004 to 2005, he was a research fellow of the Japan Society for the Promotion of Science (JSPS).
  Since 2005, he has been with the Graduate School of Informatics, Kyoto University, where he is currently an associate professor.
  From 2008 to 2009, he was a visiting researcher at Wireless@KTH, Royal Institute of Technology (KTH) in Sweden.
  He serves as an editor of IEEE Wireless Communications Letters and Journal of Communications and Information Networks, a track co-chair of APCC 2017, CCNC 2018, APCC 2018, and CCNC 2019, and a vice co-chair of IEEE ComSoc APB CCC.
  He was a tutorial lecturer in ICC 2019.
  His research interests include radio resource management, game theory, and machine learning.
  He received the PIMRC 2004 Best Student Paper Award in 2004, the Ericsson Young Scientist Award in 2006.
  He also received the Young Researcher's Award, the Paper Award, SUEMATSU-Yasuharu Award from the IEICE of Japan in 2008, 2011, and 2016, respectively, and IEEE Kansai Section GOLD Award in 2012.
  He is a senior member of the IEICE and a member of the Operations Research Society of Japan.
\end{IEEEbiography}

\begin{IEEEbiography}[{\includegraphics[width=1in,height=1.25in,clip,keepaspectratio]{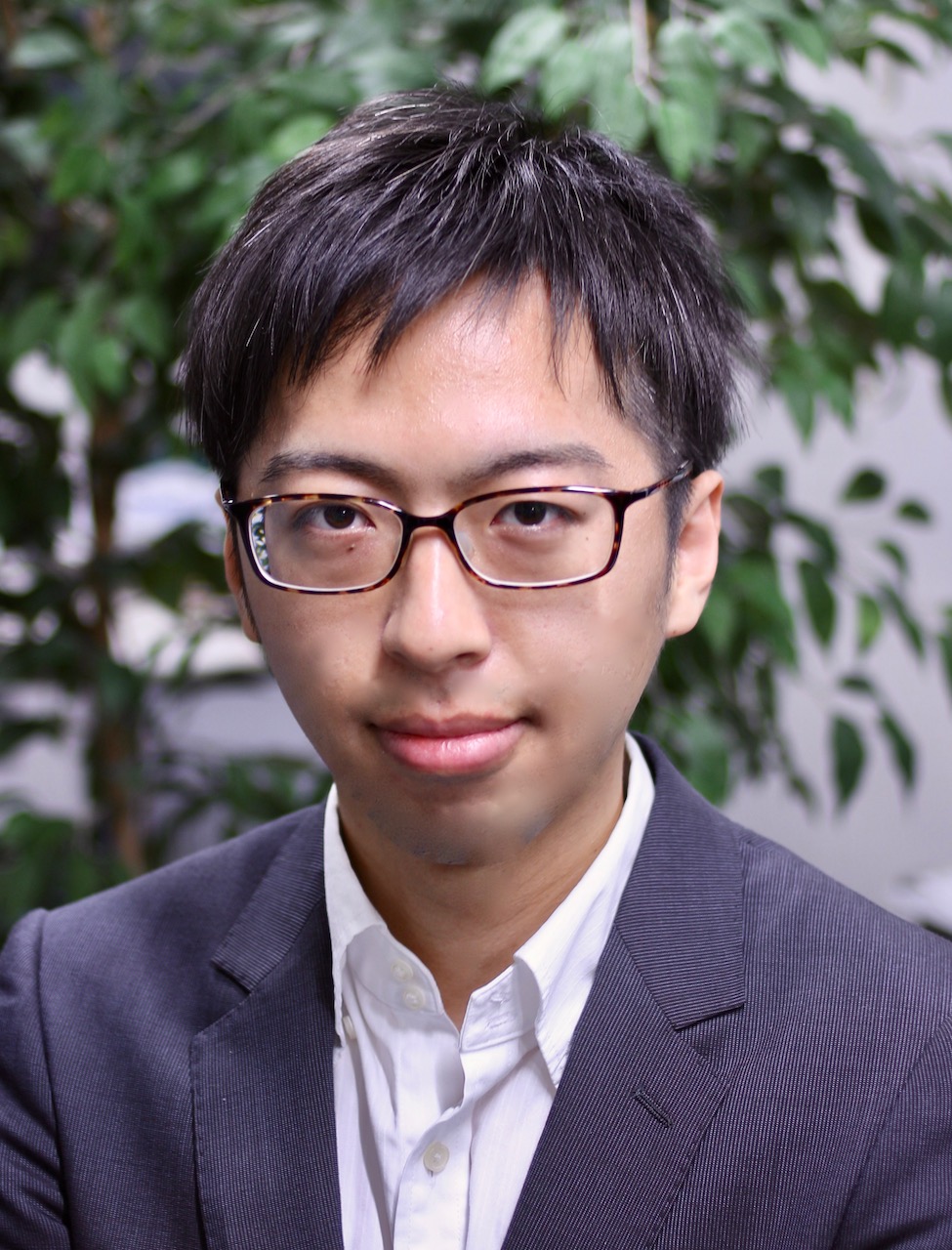}}]{Takayuki~Nishio}
  (S'11-M'14-SM'20) received the B.E.\ degree in electrical and electronic engineering and the master's and Ph.D.\ degrees in informatics from Kyoto University in 2010, 2012, and 2013, respectively. He was an assistant professor in communications and computer engineering with the Graduate School of Informatics, Kyoto University from 2013 to 2020. 
  He is currently an associate professor at the School of Engineering, Tokyo Institute of Technology, Japan.
  From 2016 to 2017, he was a visiting researcher in Wireless Information Network Laboratory (WINLAB), Rutgers University, United States. His current research interests include machine learning-based network control, machine learning in wireless networks, and heterogeneous resource management.
\end{IEEEbiography}

\begin{IEEEbiography}[{\includegraphics[width=1in,height=1.25in,clip,keepaspectratio]{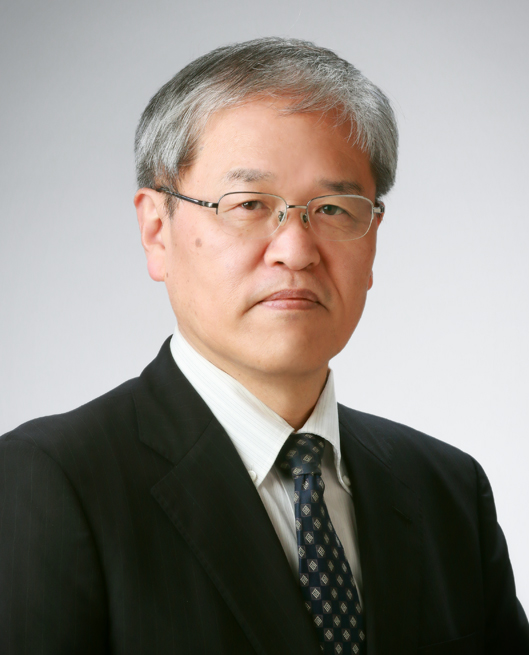}}]{Masahiro~Morikura}
  received B.E., M.E. and Ph.D. degree in electronic engineering from Kyoto University, Kyoto, Japan in 1979, 1981 and 1991, respectively. He joined NTT in 1981, where he was engaged in the research and development of TDMA equipment for satellite communications.  From 1988 to 1989, he was with the communications Research Centre, Canada as a guest scientist. From 1997 to 2002, he was active in standardization of the IEEE802.11a based wireless LAN. He received Paper Award, Achievement Award and Distinguished Achievement and Contributions Award from the IEICE in 2000, 2006 and 2019, respectively. He also received Education, Culture, Sports, Science and Technology Minister Award in 2007 and Maejima Award from the Teishin association in 2008 and the Medal of Honor with Purple Ribbon from Japan’s Cabinet Office in 2015.
  Dr. Morikura is now a professor of the Graduate School of Informatics, Kyoto University.  He is a Fellow of the IEICE and a member of IEEE.
 
\end{IEEEbiography}

\begin{IEEEbiography}[{\includegraphics[width=1in,height=1.25in,clip,keepaspectratio]{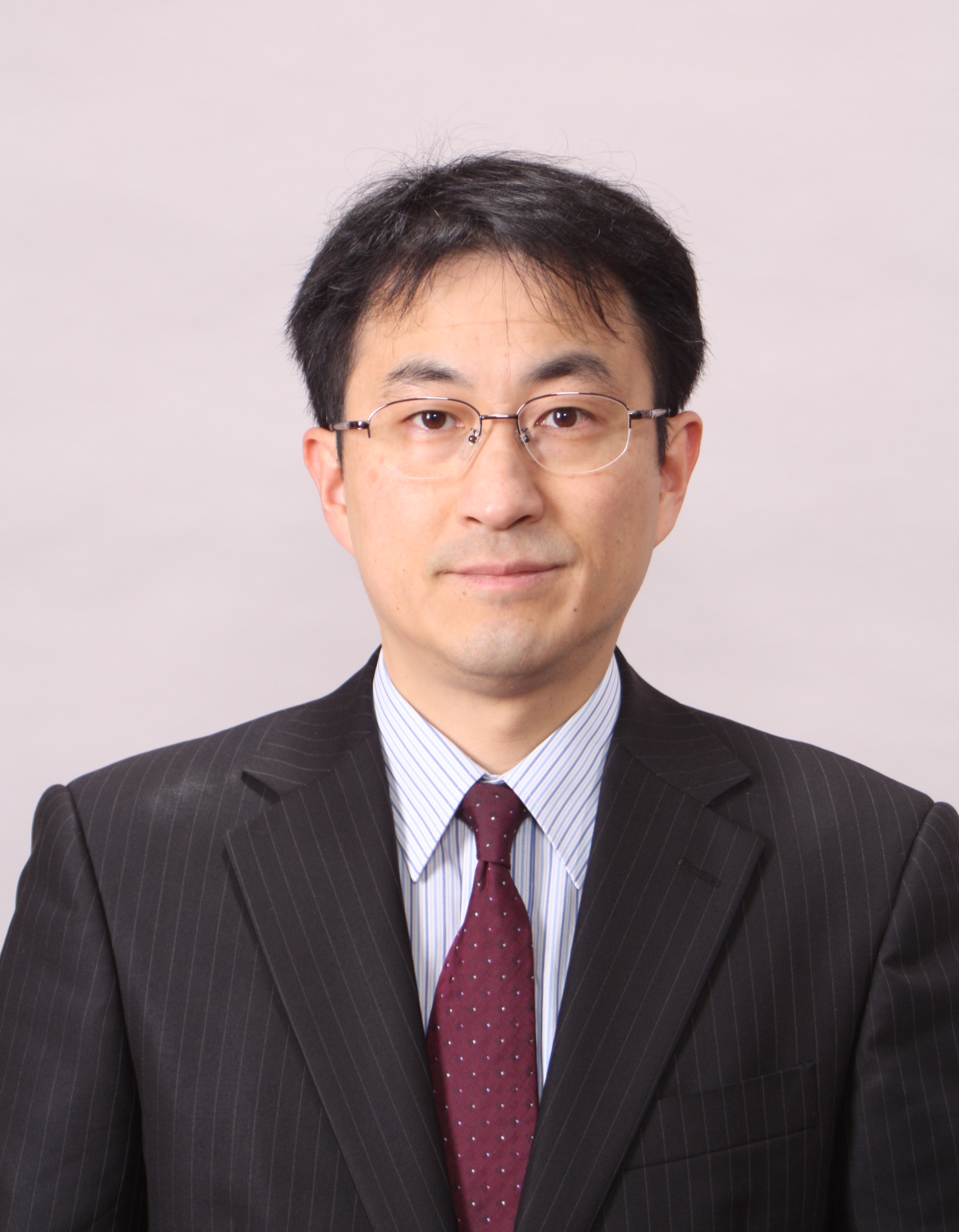}}]{Yushi Shirato}
  received B.E., M.E and D.E. degrees in electrical engineering from Tokyo University of Science in 1990, 1992 and 2018, respectively. Since joining NTT Wireless Systems Laboratories in 1992, he has been engaged in R\&D of adaptive equalizers, modems for fixed wireless access systems, and software defined radio systems. He is currently engaged in R\&D of millimeter-wave-band very high throughput fixed wireless backhaul systems. He received the Best Paper Award in 2018 and the Young Engineer’s Award in 2000 from IEICE, and the 18th Telecom System Technology Award from the Telecommunications Advancement Foundation in 2003. He is a senior member of IEICE.
\end{IEEEbiography}

\begin{IEEEbiography}[{\includegraphics[width=1in,height=1.25in,clip,keepaspectratio]{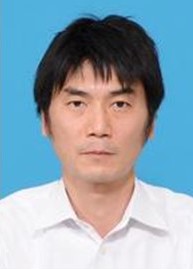}}]{Daisei Uchida}
  received B.E. and M.E. degrees in applied physics from Tokyo Institute of Technology in 1994 and 1997, respectively. Since joining NTT Wireless Systems Laboratories in 1997, he has been engaged in R\&D of the access scheme for satellite systems, radio channel assignment and tree topology of PHS-based local positioning and information systems to form wireless multi-hop networks autonomously, radio channel assignment and transmission power control for mesh-type broadband fixed wireless access systems, modulation-demodulation schemes for MIMO-OFDM systems, and modulation-demodulation schemes and radio circuit designs for Low Power Wide Area systems. He is currently engaged in R\&D of distributed antennas systems using high frequency bands for 6G generation mobile technologies. He is a member of the Institute of Electronics, Information and Communication Engineers (IEICE) of Japan. He received the Young Engineers Award from IEICE in 2001.
\end{IEEEbiography}

\begin{IEEEbiography}[{\includegraphics[width=1in,height=1.25in,clip,keepaspectratio]{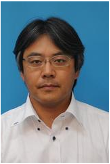}}]{Naoki Kita}
  received his B.E. degree from Tokyo Metropolitan Institute of Technology, Tokyo, Japan in 1994, and received his M.E. and Ph.D. degrees from Tokyo Institute of Technology, Tokyo, Japan in 1996 and 2007, respectively. Since joining NTT in 1996, he has been engaged in the research of propagation characteristics for wireless access systems, the development of future satellite communication systems, and international standardization on radio wave propagation. From 2009 to 2010, he was a visiting scholar at Stanford University, CA, USA. From 2013 to 2014, he was a visiting research scholar at Waseda University, Tokyo, Japan. He is currently a Senior Research Engineer, Supervisor, Group Leader at NTT Access Network Service Systems Laboratories, where he engages in the research and development of future wireless access network systems. He received the IEICE Young Reseacher’s Award, the IEICE Communications Society Best Paper Award, and IEICE Best Paper Award in 2002, 2010, and 2014, respectively. He also received the Best Paper Award in International Symposium on Antennas and Propagation 2016 (ISAP2016) in 2016. He is a senior member of IEICE and a member of IEEE. 
\end{IEEEbiography}

\end{document}